\documentclass[twocolumn,letterpaper,aps,prc,longbibliography,superscriptaddress,nofootinbib,floatfix]{revtex4-2}

\usepackage{graphicx}	
\usepackage{gensymb}
\usepackage{amsmath}

\usepackage{xspace}	

\newcommand{\mt}{\mbox{$m_T$}\xspace}
\newcommand{\pt}{\mbox{$p_T$}\xspace}
\newcommand{\pT}{\mbox{$p_T$}\xspace}
\newcommand{\ut}{\mbox{$\left<u_t\right>$}\xspace}
\newcommand{\To}{\mbox{$T_0$}\xspace}

\newcommand{\rab}{\mbox{$R_{AB}$}\xspace}
\newcommand{\Npart}{\mbox{$\langle N_{\rm part} \rangle$}\xspace}
\newcommand{\Ncoll}{\mbox{$\langle N_{\rm coll} \rangle$}\xspace}
\newcommand{\sqsn}{\mbox{$\sqrt{s_{_{NN}}}$}\xspace}
\newcommand{\pp}{\mbox{$p$$+$$p$}\xspace}
\newcommand{\dau}{\mbox{$d$$+$Au}\xspace}

\newcommand{\pal}{\mbox{$p$$+$Al}\xspace}
\newcommand{\heau}{\mbox{$^3$He$+$Au}\xspace}
\newcommand{\auau}{\mbox{Au$+$Au}\xspace}

\newcommand{\cuau}{\mbox{Cu$+$Au}\xspace}
\newcommand{\uu}{\mbox{U$+$U}\xspace}

\newcommand{\aprot}{\mbox{$\bar{p}$}\xspace}
\newcommand{\prot}{\mbox{$p$}\xspace}
\newcommand{\prots}{\mbox{$(p+\bar{p})/2$}\xspace}
\newcommand{\Km}{\mbox{$K^-$}\xspace}
\newcommand{\Kp}{\mbox{$K^+$}\xspace}
\newcommand{\Kpm}{\mbox{$K^\pm$}\xspace}
\newcommand{\pim}{\mbox{$\pi^-$}\xspace}
\newcommand{\pip}{\mbox{$\pi^+$}\xspace}
\newcommand{\pipm}{\mbox{$\pi^\pm$}\xspace}
\newcommand{\Lambd}{\mbox{$\Lambda$}\xspace}
\newcommand{\aLambd}{\mbox{$\bar{\Lambda}$}\xspace}

\begin{document}

\title{Identified charged-hadron production in $p$$+$Al, $^3$He$+$Au, and Cu$+$Au
collisions at $\sqrt{s_{_{NN}}}=200$ GeV and in U$+$U
collisions at $\sqrt{s_{_{NN}}}=193$ GeV}

\newcommand{\abilene}{Abilene Christian University, Abilene, Texas 79699, USA}
\newcommand{\augie}{Department of Physics, Augustana University, Sioux Falls, South Dakota 57197, USA}
\newcommand{\banaras}{Department of Physics, Banaras Hindu University, Varanasi 221005, India}
\newcommand{\barc}{Bhabha Atomic Research Centre, Bombay 400 085, India}
\newcommand{\baruch}{Baruch College, City University of New York, New York, New York, 10010 USA}
\newcommand{\bnlcoll}{Collider-Accelerator Department, Brookhaven National Laboratory, Upton, New York 11973-5000, USA}
\newcommand{\bnlphys}{Physics Department, Brookhaven National Laboratory, Upton, New York 11973-5000, USA}
\newcommand{\caucr}{University of California-Riverside, Riverside, California 92521, USA}
\newcommand{\charlesczech}{Charles University, Faculty of Mathematics and Physics, 180 00 Troja, Prague, Czech Republic}
\newcommand{\ciae}{Science and Technology on Nuclear Data Laboratory, China Institute of Atomic Energy, Beijing 102413, People's Republic of China}
\newcommand{\cns}{Center for Nuclear Study, Graduate School of Science, University of Tokyo, 7-3-1 Hongo, Bunkyo, Tokyo 113-0033, Japan}
\newcommand{\colorado}{University of Colorado, Boulder, Colorado 80309, USA}
\newcommand{\columbia}{Columbia University, New York, New York 10027 and Nevis Laboratories, Irvington, New York 10533, USA}
\newcommand{\czechtech}{Czech Technical University, Zikova 4, 166 36 Prague 6, Czech Republic}
\newcommand{\debrecen}{Debrecen University, H-4010 Debrecen, Egyetem t{\'e}r 1, Hungary}
\newcommand{\elte}{ELTE, E{\"o}tv{\"o}s Lor{\'a}nd University, H-1117 Budapest, P{\'a}zm{\'a}ny P.~s.~1/A, Hungary}
\newcommand{\ewha}{Ewha Womans University, Seoul 120-750, Korea}
\newcommand{\famu}{Florida A\&M University, Tallahassee, FL 32307, USA}
\newcommand{\fsu}{Florida State University, Tallahassee, Florida 32306, USA}
\newcommand{\gsu}{Georgia State University, Atlanta, Georgia 30303, USA}
\newcommand{\hanyang}{Hanyang University, Seoul 133-792, Korea}
\newcommand{\hiroshima}{Physics Program and International Institute for Sustainability with Knotted Chiral Meta Matter (SKCM2), Hiroshima University, Higashi-Hiroshima, Hiroshima 739-8526, Japan}
\newcommand{\howard}{Department of Physics and Astronomy, Howard University, Washington, DC 20059, USA}
\newcommand{\ihepprot}{IHEP Protvino, State Research Center of Russian Federation, Institute for High Energy Physics, Protvino, 142281, Russia}
\newcommand{\illuiuc}{University of Illinois at Urbana-Champaign, Urbana, Illinois 61801, USA}
\newcommand{\inrras}{Institute for Nuclear Research of the Russian Academy of Sciences, prospekt 60-letiya Oktyabrya 7a, Moscow 117312, Russia}
\newcommand{\instpasczech}{Institute of Physics, Academy of Sciences of the Czech Republic, Na Slovance 2, 182 21 Prague 8, Czech Republic}
\newcommand{\isu}{Iowa State University, Ames, Iowa 50011, USA}
\newcommand{\jaea}{Advanced Science Research Center, Japan Atomic Energy Agency, 2-4 Shirakata Shirane, Tokai-mura, Naka-gun, Ibaraki-ken 319-1195, Japan}
\newcommand{\jeonbuk}{Jeonbuk National University, Jeonju, 54896, Korea}
\newcommand{\jyvaskyla}{Helsinki Institute of Physics and University of Jyv{\"a}skyl{\"a}, P.O.Box 35, FI-40014 Jyv{\"a}skyl{\"a}, Finland}
\newcommand{\kek}{KEK, High Energy Accelerator Research Organization, Tsukuba, Ibaraki 305-0801, Japan}
\newcommand{\korea}{Korea University, Seoul 02841, Korea}
\newcommand{\kurchatov}{National Research Center ``Kurchatov Institute", Moscow, 123098 Russia}
\newcommand{\kyoto}{Kyoto University, Kyoto 606-8502, Japan}
\newcommand{\labllr}{Laboratoire Leprince-Ringuet, Ecole Polytechnique, CNRS-IN2P3, Route de Saclay, F-91128, Palaiseau, France}
\newcommand{\lahorelums}{Physics Department, Lahore University of Management Sciences, Lahore 54792, Pakistan}
\newcommand{\lawllnl}{Lawrence Livermore National Laboratory, Livermore, California 94550, USA}
\newcommand{\losalamos}{Los Alamos National Laboratory, Los Alamos, New Mexico 87545, USA}
\newcommand{\lund}{Department of Physics, Lund University, Box 118, SE-221 00 Lund, Sweden}
\newcommand{\lyon}{IPNL, CNRS/IN2P3, Univ Lyon, Universit{\'e} Lyon 1, F-69622, Villeurbanne, France}
\newcommand{\maryland}{University of Maryland, College Park, Maryland 20742, USA}
\newcommand{\mass}{Department of Physics, University of Massachusetts, Amherst, Massachusetts 01003-9337, USA}
\newcommand{\mate}{MATE, Laboratory of Femtoscopy, K\'aroly R\'obert Campus, H-3200 Gy\"ongy\"os, M\'atrai\'ut 36, Hungary}
\newcommand{\michigan}{Department of Physics, University of Michigan, Ann Arbor, Michigan 48109-1040, USA}
\newcommand{\miss}{Mississippi State University, Mississippi State, Mississippi 39762, USA}
\newcommand{\muhlenberg}{Muhlenberg College, Allentown, Pennsylvania 18104-5586, USA}
\newcommand{\myongji}{Myongji University, Yongin, Kyonggido 449-728, Korea}
\newcommand{\nagasaki}{Nagasaki Institute of Applied Science, Nagasaki-shi, Nagasaki 851-0193, Japan}
\newcommand{\nara}{Nara Women's University, Kita-uoya Nishi-machi Nara 630-8506, Japan}
\newcommand{\natmephi}{National Research Nuclear University, MEPhI, Moscow Engineering Physics Institute, Moscow, 115409, Russia}
\newcommand{\newmex}{University of New Mexico, Albuquerque, New Mexico 87131, USA}
\newcommand{\nmsu}{New Mexico State University, Las Cruces, New Mexico 88003, USA}
\newcommand{\northcg}{Physics and Astronomy Department, University of North Carolina at Greensboro, Greensboro, North Carolina 27412, USA}
\newcommand{\ohio}{Department of Physics and Astronomy, Ohio University, Athens, Ohio 45701, USA}
\newcommand{\ornl}{Oak Ridge National Laboratory, Oak Ridge, Tennessee 37831, USA}
\newcommand{\orsay}{IPN-Orsay, Univ.~Paris-Sud, CNRS/IN2P3, Universit\'e Paris-Saclay, BP1, F-91406, Orsay, France}
\newcommand{\peking}{Peking University, Beijing 100871, People's Republic of China}
\newcommand{\pnpi}{PNPI, Petersburg Nuclear Physics Institute, Gatchina, Leningrad region, 188300, Russia}
\newcommand{\pusan}{Pusan National University, Pusan 46241, Korea}
\newcommand{\riken}{RIKEN Nishina Center for Accelerator-Based Science, Wako, Saitama 351-0198, Japan}
\newcommand{\rikjrbrc}{RIKEN BNL Research Center, Brookhaven National Laboratory, Upton, New York 11973-5000, USA}
\newcommand{\rikkyo}{Physics Department, Rikkyo University, 3-34-1 Nishi-Ikebukuro, Toshima, Tokyo 171-8501, Japan}
\newcommand{\saispbstu}{Saint Petersburg State Polytechnic University, St.~Petersburg, 195251 Russia}
\newcommand{\seoulnat}{Department of Physics and Astronomy, Seoul National University, Seoul 151-742, Korea}
\newcommand{\stonybrkc}{Chemistry Department, Stony Brook University, SUNY, Stony Brook, New York 11794-3400, USA}
\newcommand{\stonycrkp}{Department of Physics and Astronomy, Stony Brook University, SUNY, Stony Brook, New York 11794-3800, USA}
\newcommand{\sungskku}{Sungkyunkwan University, Suwon, 440-746, Korea}
\newcommand{\tenn}{University of Tennessee, Knoxville, Tennessee 37996, USA}
\newcommand{\texsu}{Texas Southern University, Houston, TX 77004, USA}
\newcommand{\titech}{Department of Physics, Tokyo Institute of Technology, Oh-okayama, Meguro, Tokyo 152-8551, Japan}
\newcommand{\tsukuba}{Tomonaga Center for the History of the Universe, University of Tsukuba, Tsukuba, Ibaraki 305, Japan}
\newcommand{\vandy}{Vanderbilt University, Nashville, Tennessee 37235, USA}
\newcommand{\weizmann}{Weizmann Institute, Rehovot 76100, Israel}
\newcommand{\wigner}{Institute for Particle and Nuclear Physics, Wigner Research Centre for Physics, Hungarian Academy of Sciences (Wigner RCP, RMKI) H-1525 Budapest 114, POBox 49, Budapest, Hungary}
\newcommand{\yonsei}{Yonsei University, IPAP, Seoul 120-749, Korea}
\newcommand{\zagreb}{Department of Physics, Faculty of Science, University of Zagreb, Bijeni\v{c}ka c.~32 HR-10002 Zagreb, Croatia}
\newcommand{\zambia}{Department of Physics, School of Natural Sciences, University of Zambia, Great East Road Campus, Box 32379, Lusaka, Zambia}
\affiliation{\abilene}
\affiliation{\augie}
\affiliation{\banaras}
\affiliation{\barc}
\affiliation{\baruch}
\affiliation{\bnlcoll}
\affiliation{\bnlphys}
\affiliation{\caucr}
\affiliation{\charlesczech}
\affiliation{\ciae}
\affiliation{\cns}
\affiliation{\colorado}
\affiliation{\columbia}
\affiliation{\czechtech}
\affiliation{\debrecen}
\affiliation{\elte}
\affiliation{\ewha}
\affiliation{\famu}
\affiliation{\fsu}
\affiliation{\gsu}
\affiliation{\hanyang}
\affiliation{\hiroshima}
\affiliation{\howard}
\affiliation{\ihepprot}
\affiliation{\illuiuc}
\affiliation{\inrras}
\affiliation{\instpasczech}
\affiliation{\isu}
\affiliation{\jaea}
\affiliation{\jeonbuk}
\affiliation{\jyvaskyla}
\affiliation{\kek}
\affiliation{\korea}
\affiliation{\kurchatov}
\affiliation{\kyoto}
\affiliation{\labllr}
\affiliation{\lahorelums}
\affiliation{\lawllnl}
\affiliation{\losalamos}
\affiliation{\lund}
\affiliation{\lyon}
\affiliation{\maryland}
\affiliation{\mass}
\affiliation{\mate}
\affiliation{\michigan}
\affiliation{\miss}
\affiliation{\muhlenberg}
\affiliation{\myongji}
\affiliation{\nagasaki}
\affiliation{\nara}
\affiliation{\natmephi}
\affiliation{\newmex}
\affiliation{\nmsu}
\affiliation{\northcg}
\affiliation{\ohio}
\affiliation{\ornl}
\affiliation{\orsay}
\affiliation{\peking}
\affiliation{\pnpi}
\affiliation{\pusan}
\affiliation{\riken}
\affiliation{\rikjrbrc}
\affiliation{\rikkyo}
\affiliation{\saispbstu}
\affiliation{\seoulnat}
\affiliation{\stonybrkc}
\affiliation{\stonycrkp}
\affiliation{\sungskku}
\affiliation{\tenn}
\affiliation{\texsu}
\affiliation{\titech}
\affiliation{\tsukuba}
\affiliation{\vandy}
\affiliation{\weizmann}
\affiliation{\wigner}
\affiliation{\yonsei}
\affiliation{\zagreb}
\affiliation{\zambia}
\author{N.J.~Abdulameer} \affiliation{\debrecen}
\author{U.~Acharya} \affiliation{\gsu} 
\author{A.~Adare} \affiliation{\colorado} 
\author{C.~Aidala} \affiliation{\losalamos} \affiliation{\michigan} 
\author{N.N.~Ajitanand} \altaffiliation{Deceased} \affiliation{\stonybrkc} 
\author{Y.~Akiba} \email[PHENIX Spokesperson: ]{akiba@rcf.rhic.bnl.gov} \affiliation{\riken} \affiliation{\rikjrbrc} 
\author{R.~Akimoto} \affiliation{\cns} 
\author{J.~Alexander} \affiliation{\stonybrkc} 
\author{M.~Alfred} \affiliation{\howard} 
\author{V.~Andrieux} \affiliation{\michigan} 
% ...
\collaboration{PHENIX Collaboration}  \noaffiliation

\date{\today}


\begin{abstract}


The PHENIX experiment has performed a systematic study of identified
charged-hadron ($\pi^\pm$, $K^\pm$, $p$, $\bar{p}$) production at
midrapidity in $p$$+$Al, $^3$He$+$Au, Cu$+$Au collisions at
$\sqrt{s_{_{NN}}}=200$ GeV and U$+$U collisions at
$\sqrt{s_{_{NN}}}=193$ GeV. Identified charged-hadron invariant
transverse-momentum ($p_T$) and transverse-mass ($m_T$) spectra are
presented and interpreted in terms of radially expanding thermalized
systems. The particle ratios of $K/\pi$ and $p/\pi$ have been measured
in different centrality ranges of large (Cu$+$Au, U$+$U) and small
($p$$+$Al, $^3$He$+$Au) collision systems. The values of $K/\pi$ ratios
measured in all considered collision systems were found to be consistent
with those measured in $p$$+$$p$ collisions. However the values of
$p/\pi$ ratios measured in large collision systems reach the values of
$\approx\,0.6$, which is $\approx\,2$ times larger than in $p$$+$$p$
collisions. These results can be qualitatively understood in terms of
the baryon enhancement expected from hadronization by recombination.
Identified charged-hadron nuclear-modification factors ($R_{AB}$) are
also presented. Enhancement of proton $R_{AB}$ values over meson
$R_{AB}$ values was observed in central $^3$He$+$Au, Cu$+$Au, and U$+$U
collisions. The proton $R_{AB}$ values measured in $p$$+$Al collision
system were found to be consistent with $R_{AB}$ values of $\phi$,
$\pi^\pm$, $K^\pm$, and $\pi^0$ mesons, which may indicate that the size
of the system produced in $p$$+$Al collisions is too small for
recombination to cause a noticeable increase in proton production.

\end{abstract}

\maketitle

\section{INTRODUCTION}

The quark-gluon plasma (QGP)~\cite{QGP,QGP_STAR} is a state of matter 
that exists at extremely high temperature and density. The QGP comprises 
deconfined, strongly interacting quarks and gluons, which are ordinarily 
confined inside hadrons. Quantum chromodynamics, the theory of the 
strong nuclear force, predicts QGP formation in high-energy collisions 
of heavy nuclei. The QGP existence have been verified by experimental 
observations of QGP signatures~\cite{QGP,QGP_signatures}, such as 
strangeness 
enhancement~\cite{StrangenessEnhancenment,Strangeness_QGP,phi_LargeSysts,ALICE_strangeness,STAR_strangeness,BRAHMS_Strangeness_CH}, 
jet quenching~\cite{JetQuenching1,JetQuenching2,JetQuenching3,ALICE_jet_quenching}, 
and baryon 
enhancement~\cite{PPG026,CHspectra_130GeV,Velkovska_2002,ppg146,PHOBOS_CH,STAR_CH}.

The QGP is formed in heavy ion collisions (such as \cuau, \uu) at 
temperatures larger than 300~MeV~\cite{Coalescence_models,STAR_UU}, but 
expands and cools down in a short time 
($\approx\,10$~fm/$c$~\cite{QGP_signatures}). When the critical temperature 
($\approx\,175$~MeV~\cite{QGP,Coalescence_models}) is reached, the 
hadronization process begins. It was expected~\cite{Coalescence_models} 
that in relativistic ion collisions protons are produced roughly 3 times 
less than pions, reflecting the mass difference between them and the 
requirement for a nonzero baryon number to form a proton. Indeed, the 
values of the proton to pion ratio ($p/\pi$) measured by the PHENIX 
experiment in \pp collisions do not exceed a value of 0.3. However, in 
central Au$+$Au collisions, the $p/\pi$ ratio reaches a value of 0.8, 
indicating that protons and $\pi$-mesons are produced in nearly equal 
proportion. The anomalous proton production observed by the PHENIX 
experiment~\cite{PHENIXoverview} in Au$+$Au collisions at 
\sqsn=130~GeV~\cite{CHspectra_130GeV} and later at 
\sqsn=200~GeV~\cite{Velkovska_2002} was successfully described by the 
recombination model. The recombination 
model~\cite{Recombination1,Recombination2} is a hadronization model that 
considers hadron formation as a result of combining quarks that are 
nearby in phase space.

Small collision systems (such as \pal, \heau) were previously 
studied primarily to investigate cold-nuclear-matter 
effects~\cite{CNM,phi_dAu} (Cronin 
enhancement~\cite{Cronin,Cronin_STAR}, multiple-parton 
scattering~\cite{MPI2}, modifications of the initial nuclear-parton distribution 
functions~\cite{PDF1,PDF2}). It was thought~\cite{PHENIX_Nature,CNM} 
that the size and lifetime of the system produced in small collision 
systems are not sufficient for QGP formation.  However, in 2019 the 
PHENIX experiment reported on the observation of elliptic and triangular 
flow patterns of charged particles produced in $p$/$d$/\heau 
collisions~\cite{PHENIX_Nature}. The results of these measurements have 
been described in the frame of hydrodynamical models, which include the 
formation of short-lived QGP droplets.
Comparison of identified charged-hadron production in small and large 
collision systems may enable systematic studies to determine the minimal 
conditions of QGP formation and to investigate influence of collision 
geometry and system size on hadron production.

This paper presents new data on identified charged-hadron production in 
\pal, \heau, Cu$+$Au collisions at \sqsn~=~200 GeV and U$+$U collisions 
at \sqsn~=~193~GeV.  Discussed are the influence of collision centrality 
and geometry on identified charged-hadron invariant \pt and \mt spectra, 
nuclear-modification factors, and ratios.


\section{DATA ANALYSIS}

The data sets used in the present analysis were collected by the PHENIX 
experiment during the operational periods of the Relativistic Heavy Ion 
Collider in calendar years 2012 (Cu$+$Au and U$+$U), 2014 (\heau), and 
2015 (\pal). The PHENIX experiment~\cite{PHENIXoverview} comprised 
several subdetectors that are grouped into four main blocks called arms. 
The central arms---east and west---cover midrapidity ($|\eta|<0.35$) and 
are intended for measurement of electrons, photons, and charged hadrons. 
In the present analysis, the drift chambers (DC)~\cite{TrackingSystem} 
are used for momentum determination, the time-of-flight (TOF) 
wall~\cite{ToF} is used for charged hadron identification, and the 
beam-beam counters (BBC)~\cite{PHENIX_InnerDetectors} are used for 
centrality determination. Detailed information about the PHENIX 
experiment can be found in 
Refs.~\cite{PHENIXoverview,PHENIX_CAdetectors,PHENIX_InnerDetectors}.

\subsection{Event selection}

The minimum-bias (MB) trigger was used for event selection. The MB 
trigger selects events that correspond to simultaneous signals in the 
north (forward rapidity) and south (backward rapidity) BBC detectors. 
The collision vertex coordinate is required to be 
$|z_{{\rm vtx}}|<30~{\rm cm}$ relative to the center of the 
spectrometer. Determination of centrality in the PHENIX experiment is 
done with the BBC according to the procedure described 
in~\cite{CentralityIdentification}. The numbers of participating 
nucleons \Npart as well as numbers of binary nucleon-nucleon 
collisions \Ncoll are calculated using Glauber Monte Carlo 
simulations. The \Npart and \Ncoll values are presented in the 
Table~\ref{table:MC_parameters}.

\begin{table}[!htb]
\begin{minipage}{0.995\linewidth}
\caption{\label{table:MC_parameters}
Summary of the \Ncoll and \Npart values calculated using Glauber Monte Carlo 
simulation.}
 \begin{ruledtabular}
 \begin{tabular}{ccccc}
 System  & Centrality & \Ncoll    &  \Npart       \\ \hline
\pal & 0\%--72\%     & 2.1$\pm$0.2    & 3.1$\pm$0.1    \\
     & 0\%--20\%     & 3.4$\pm$0.3    & 4.4$\pm$0.3    \\
     &20\%--40\%    & 2.3$\pm$0.2    & 3.3$\pm$0.1    \\
     &40\%--72\%    & 1.7$\pm$0.1   & 1.6$\pm$0.2  \\ 
\\
\heau &  0\%--88\%     & 10.4$\pm$0.7 & 11.3$\pm$0.5    \\
      &  0\%--20\%     & 22.3$\pm$1.7 & 21.1$\pm$1.3    \\
      & 20\%--40\%    & 14.8$\pm$1.1 & 15.4$\pm$0.9    \\
      & 40\%--60\%    & 8.4$\pm$0.6  & 9.5$\pm$0.6     \\
      &  0\%--88\%    & 3.4$\pm$0.3  & 4.8$\pm$0.3     \\ 
\\  
\cuau &  0\%--80\%     & 123.8$\pm$12.0  & 70.4$\pm$3.0    \\
      &  0\%--20\%     & 313.8$\pm$28.4  & 154.8$\pm$4.1     \\
      & 20\%--40\%    & 129.3$\pm$12.4  & 80.4$\pm$3.3     \\
      & 40\%--60\%    & 41.8$\pm$5.3    & 34.9$\pm$2.9   \\
      & 60\%--80\%    & 10.1$\pm$2.0    & 11.5$\pm$1.8   \\ 
\\ 
\uu &  0\%--80\%     & 342$\pm$30    & 143$\pm$5   \\
    &  0\%--20\%     & 935$\pm$98    & 330$\pm$6   \\
    & 20\%--40\%    & 335$\pm$33    & 259$\pm$7   \\
    & 40\%--60\%    & 81$\pm$13     & 65$\pm$6    \\
    & 60\%--80\%    & 17$\pm$4      & 18$\pm$3   \\
 \end{tabular}
 \end{ruledtabular}
\end{minipage}
\vspace{0.5cm}
\begin{minipage}{0.995\linewidth}
 \caption{\label{table:pt_ranges}
The \pt ranges (GeV/$c$) of identified charged-hadron yields 
measurements.}
 \begin{ruledtabular} \begin{tabular}{ccccc}
Hadron &  \pal & \heau  & Cu$+$Au & U$+$U  \\ \hline
\pip      & 0.5--2.0  &  0.5--3.0  &  0.5--3.0  & 0.5--3.0  \\
\pim      & 0.5--2.0  &  0.5--3.0  &  0.5--3.0  & 0.5--3.0  \\
\Kp       & 0.5--1.8  &  0.5--2.0  &  0.5--2.0  & 0.5--2.0  \\
\Km       & 0.5--1.8  &  0.5--2.0  &  0.5--2.0  & 0.5--2.0  \\
\prot     &           &  0.5--4.0  &  0.5--4.0  & 0.5--4.0  \\
\aprot    & 0.5--2.5  &  0.5--4.0  &  0.5--4.0  & 0.5--4.0  \\
\end{tabular} \end{ruledtabular}
\end{minipage}
 \end{table}

\subsection{Particle identification}


Charged hadrons are detected in the TOF wall and the DC of the PHENIX 
experiment. The squared mass of tracks can be determined in accordance 
with Eq.~\ref{eq:TOFm2}:
\begin{equation}
    \label{eq:TOFm2}
    m^2 = \frac{p^2}{c^2} \left(\frac{t^2 c^2}{L^2} - 1 \right),
\end{equation}
where $p$ is the momentum of the particle measured with DC, $L$ is the 
particle flight path-length from the event vertex to the TOF detector, 
$t$ is the time-of-flight, measured in TOF detector, $c$ is the speed of 
light~\cite{PHENIX_CAdetectors}. Table~\ref{table:pt_ranges} presents 
\pt ranges of identified charged-hadron-yield measurements.

\begin{figure}[!tbh]
\includegraphics[width=1.0\linewidth]{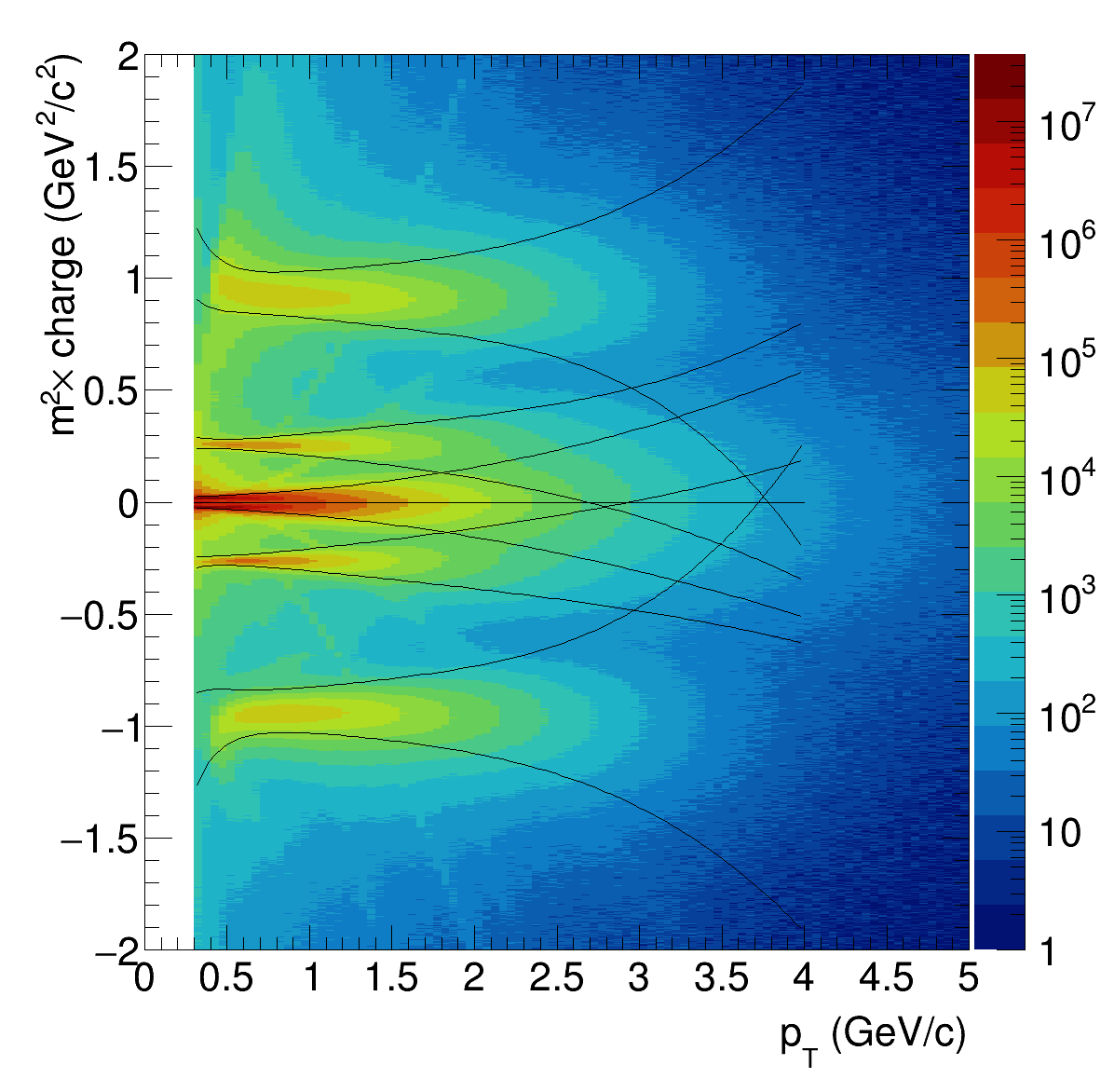}
\caption{Distribution of hadron squared mass multiplied by charge 
($m^{2} \cdot {\rm charge}$) vs.~hadron \pT as found from TOF timing, DC 
momentum, and the path length. Black solid lines represent PID cuts, 
based on Eq.~\ref{eq:TOFm2gaus_approx}, which were used for hadron 
identification.}
\label{fig:m2pT}
\end{figure}

The distribution of $m^2$ multiplied by charge versus transverse 
momentum (\pt) as found from TOF timing, DC momentum, and the path 
length is presented in Fig.~\ref{fig:m2pT}. Signals corresponding to 
$\pi^{\pm}$, $K^{\pm}$, and $p$ and $\bar{p}$ are clearly 
distinguishable. In \pal collisions, the proton sample has a large 
contamination of spallation protons from the inner silicon detectors. 
For that reason, they are excluded from measurements that require 
absolute normalization, like the spectra, ratios, and nuclear 
modification factors reported in this manuscript.

Particle identification (PID) was carried out by applying 
two-standard-deviation (2$\sigma$) PID cuts in $m^2$ and momentum space 
for each particle species. The hadron's signals from TOF are 
approximated by Gaussian functions with the root-mean-square deviations 
($\sigma_{\rm TOF}$) and mathematical expectations ($m_{\rm TOF}^2$) in 
every $\Delta p_T = 0.5$~GeV/$c$ interval of the hadron identification 
\pt range from Table~\ref{table:pt_ranges}. Discrete $\sigma_{\rm 
TOF}(p_T)$ and $m_{\rm TOF}^2(p_T)$ dependencies are parameterized by 
Eq.~\ref{eq:TOFm2gaus_approx}. PID cuts are based on obtained 
continuous functions $\sigma_{\rm TOF}(p_T)$ and $m_{\rm TOF}^2(p_T)$, 
which are presented in Fig.~\ref{fig:m2pT} with black solid lines. The 
fit to both the mean $m_{\rm TOF}^2(p_T)$ and the standard deviation 
$\sigma_{\rm TOF}(p_T)$ uses the same functional form,

\begin{equation}
    \label{eq:TOFm2gaus_approx}
    m_{\rm TOF}^2(p_T),\sigma_{\rm TOF}(p_T) = p_0 +\frac{p_1}{p_T} + \frac{p_2}{p_T^2} + p_3 \cdot e^{\sqrt{p_T}} +p_4 \cdot \sqrt{p_T},
\end{equation}
where $p_0$, $p_1$, $p_2$, $p_3$, and $p_4$ are fit parameters,
and the set of fit parameters is different for
$m_{\rm TOF}^2(p_T)$ and $\sigma_{\rm TOF}(p_T)$.
In the high-$p_T$ region, hadron signals start to
overlap; therefore, the veto cut was introduced for better separation of
$\pi$, $K$ and $p$.  The veto cut requires that the hadron mass does not
satisfy the $1.5\,\sigma$ condition for neighboring hadrons. The PID and
veto cuts are standard for the PHENIX detector~\cite{PPG026,ppg146}.

\subsection{Corrections to the raw data}

The measured values of identified charged hadron primary yields should 
be corrected for the geometric acceptance of the detectors, detector 
efficiency, and analysis cuts used in data 
selection~\cite{PPG026,ppg146}. This section describes procedures of 
estimating the corrections applied to the raw identified charged-hadron 
yields.

\subsubsection{Reconstruction efficiency}

\begin{figure}[!tbh]
\includegraphics[width=1.0\linewidth]{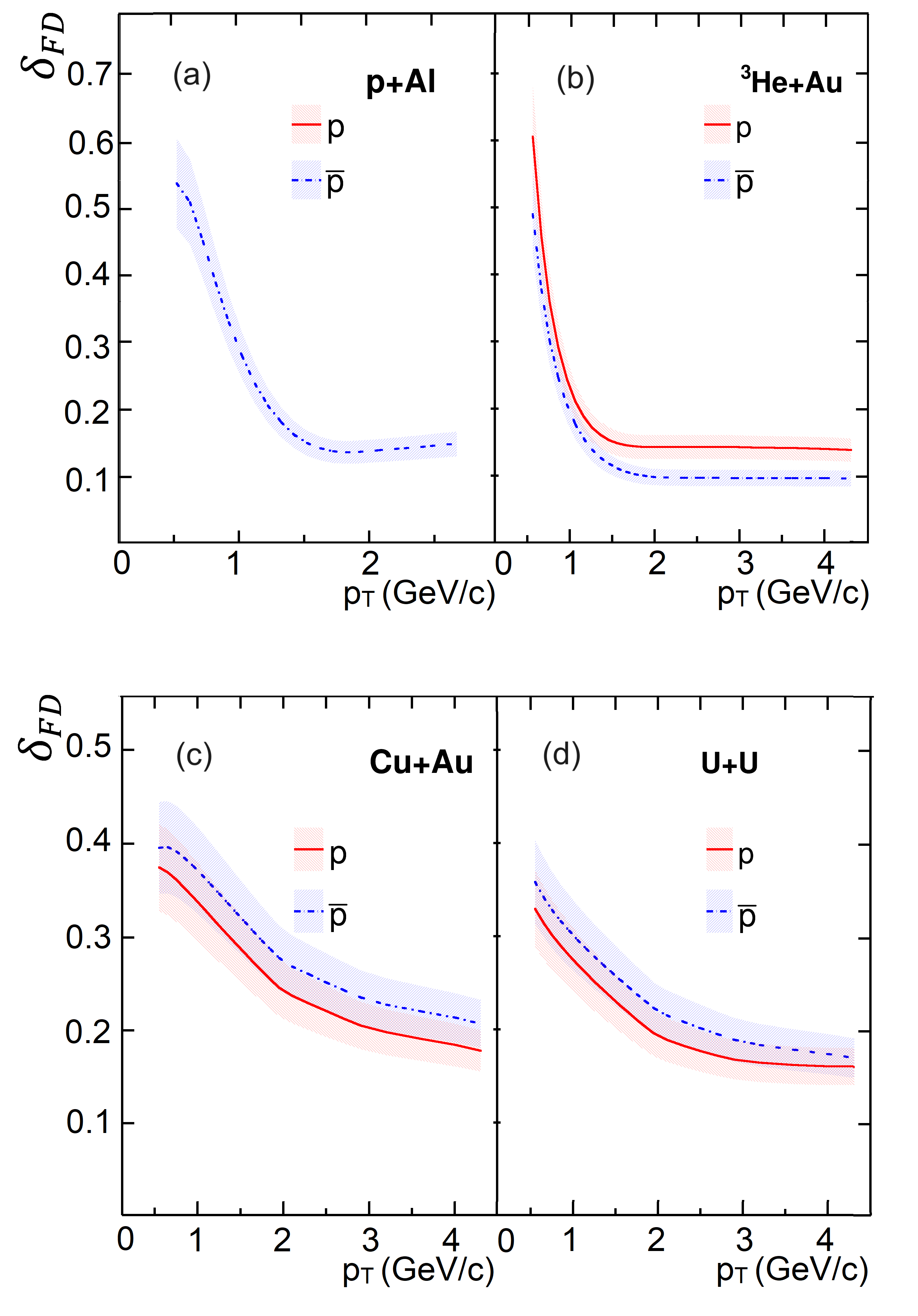}
\caption{$\delta_{\rm FD}$ values as a function of \pt for protons and 
antiprotons, measured in \pal, \heau, Cu$+$Au, and U$+$U collisions. The 
shaded bands indicate the systematic uncertainty.}
\label{fig:FeedDown}
\end{figure}


The geometric acceptance of the detectors, detector efficiency, and 
analysis cuts have been taken into account by the reconstruction 
efficiency factor ($\epsilon_{\rm rec}$), estimated using a single-particle 
Monte-Carlo (MC) simulation. The MC simulation was carried out using the 
PHENIX integrated simulation application (PISA) package~\cite{PISA}, 
which is based on {\sc geant}3~\cite{GEANT} simulation software. In the 
PISA project the geometry, material, spatial, momentum and energy 
resolutions of the PHENIX detector subsystems as well as the 
configuration of the magnetic field are simulated in full accordance 
with the structure of the real facility. Simulated events have been 
processed using the same analysis criteria as the data. In that way, the 
reconstruction efficiency $\epsilon_{\rm rec}$(\pT) can be estimated by 
the number of reconstructed hadrons ($N_{\rm rec}^{\rm MC}$) divided by 
the total number of hadrons generated in the MC model ($N_{\rm tot}^{\rm 
MC}$):

\begin{equation}
  \label{eq:CorrFactor}
    \epsilon_{\rm rec}(p_T) = \frac{dN_{\rm rec}^{\rm MC}/dp_T}{dN_{\rm tot}^{\rm MC}/dp_T}.
\end{equation}

\subsubsection{Weak-decay correction}

Hyperon weak decays are an additional source of contamination for \prot 
and \aprot yields. The fraction of detected \prot and \aprot that come 
from hyperon weak decays can be taken into account by feed-down 
corrections $C_{\rm FD}=1-\delta_{\rm FD}$, where $\delta_{\rm FD}$ is 
determined as a fraction of \prot (or \aprot) that come from hyperon 
weak decays to the total number of registered protons (antiprotons). The 
greatest contribution to the amount of decay protons and antiprotons is 
made by \Lambd and \aLambd decays~\cite{PPG026}. The contribution of 
$\Sigma^0$ is implicitly taken into account when considering \Lambd and 
\aLambd decays, because $\Sigma^0$ decays electromagnetically with 100\% 
branching ratio to \Lambd and a photon. Other weak hyperon decays, such 
as $\Sigma^{\pm}$, $\Xi$, etc. have not been considered in the present 
analyzes.  The contribution of charged $\Sigma$ states, $\Xi$ and 
$\Omega$ multistrange baryon states is 
small~\cite{ppg146,CentralityIdentification} and was included in the 
overall systematic uncertainty estimates.

MC simulations of \Lambd and \aLambd are used to calculate the fraction 
of \prot and \aprot that come from \Lambd and \aLambd decays. 
Figure~\ref{fig:FeedDown} shows the values of $\delta_{\rm FD}$ as a 
function of \pt, obtained in \pal, \heau, Cu$+$Au, and U$+$U collisions.

\subsubsection{Bias-factor correction}

The MB trigger is biased for events with high multiplicity, leading to 
miscalculations of hadron invariant yields in small collision 
systems~\cite{BiasFactor}. The bias of the MB trigger arises mostly from 
single- and double-diffractive events, which result in particle 
production very close to the beam line. The BBC trigger is therefore 
biased to the nondiffractive collisions, which have larger particle 
production at midrapidity. In such cases, there will exist a bias 
towards higher charge deposition in the BBC and, hence, towards larger 
centrality~\cite{BiasFactor}. The bias effect was corrected by 
applying bias-factor corrections ($f_{\rm bias}$). The values of 
$f_{\rm bias}$ corrections are estimated with the help of Glauber MC 
simulations and presented in Table~\ref{table:BiasFactor}.

\subsection{Systematic uncertainties}

Systematic uncertainties are divided into three types: type A, type B, 
and type C. Uncertainties of type A are point-to-point uncorrelated in 
transverse momentum and centrality. Type B uncertainties are 
point-to-point correlated with transverse momentum. Type A and B 
uncertainties arise as a result of acceptance uncertainties, weak decay 
correction uncertainty, applying track selection and PID, cuts. The 
values of type A and B uncertainties have been estimated by varying 
analysis-cut parameters. Track selection and acceptance uncertainties 
mostly cancel for measurements of $p/\pi$ and $K/\pi$ 
ratios~\cite{ppg146,PPG026}, so only uncertainties associated with PID cuts and 
weak decay correction are shown with the results presented in 
Table~\ref{table:syst}.


Uncertainties of type C are fully correlated in transverse momentum, 
i.e.~uncertainties of type C shift hadron yields equally in the entire 
range of the transverse momentum. In this work, the Type C uncertainties 
are attributed to the uncertainty in estimation of \Ncoll and 
$f_{\rm bias}$ values.


\begingroup \squeezetable
\begin{table}[!tbh]
 \caption{\label{table:BiasFactor}
Summary of the $f_{\rm bias}$ correction values calculated using Glauber 
Monte Carlo simulation.}
 \begin{ruledtabular}
 \begin{tabular}{ccc}
Collision system & Centrality &  $f_{\rm bias}$  \\ 
\hline
\pal & 0\%--72\%     & 0.80$\pm$0.02       \\
     & 0\%--20\%     & 0.81$\pm$0.01       \\
     & 20\%--40\%    & 0.90$\pm$0.02       \\
     & 40\%--72\%    & 1.05$\pm$0.04    \\ 
\\
\heau & 0\%--88\%     & 0.89$\pm$0.01  \\
      & 0\%--20\%     & 0.95$\pm$0.01  \\
      & 20\%--40\%    & 1.01$\pm$0.01  \\
      & 40\%--60\%    & 1.02$\pm$0.01   \\
      & 60\%--88\%    & 1.03$\pm$0.05   \\
 \end{tabular} \end{ruledtabular}
\end{table}

\begin{table}[tbh]
\caption{\label{table:syst}
Values of type A uncertainties summed quadratically to the 
values of type B uncertainties (\%) on the identified charged hadron 
invariant yields and ratios. The \pt ranges correspond to 
Table~\ref{table:pt_ranges}}.
\vspace{-0.2cm}
 \begin{ruledtabular}
\begin{tabular}{ccccccc}
&     &    &    & $p_{T}$ (GeV/$c$) &  & \\
System & Hadron   &  0.5--1 & 1--1.5  & 1.5--2.0 & 2.0\%--3.0 & 3.0\%--4.0   \\ 
\hline
\pal & \pip  &      9.7  &      10.5  &      11.3  &       &           \\
     & \pim  &      8.7  &      11.0  &      10.5  &         &           \\
     & \Kp  &      7.9  &      10.2  &      13.7  &         &           \\
     & \Km  &      7.7  &      9.9  &      15.0  &         &           \\
     & \aprot  &      9.3  &      7.9  &      7.7  &      8.8  &          \\
     & $K^{+}/\pi^{+}$  &  3.6 &  3.9 &    8.1 &     &          \\
     & $K^{-}/\pi^{-}$  &  3.6 &  3.7 &    8.7 &     &          \\
     & $\bar{p}/\pi^{-}$  &  6.0 &  6.1 &   6.6 &   6.6  &         \\ \\

\heau & \pip  &      5.7  &      4.2  &      5.7  &      6.7  &           \\
     & \pim  &      11.6  &      11.3  &      10.5  &      9.1  &           \\
     & \Kp  &      8.3  &      8.6  &      10.0  &         &           \\
     & \Km  &      8.7  &      9.9  &      11.3  &         &           \\
     & \prot  &      8.6  &      8.6  &      8.8  &      8.5  &      8.8    \\
     & \aprot  &      9.1  &      10.0  &      10.4  &      10.3  &      10.6    \\ 
     & $K^{+}/\pi^{+}$  &  5.1 &  4.0 &    5.1 &     &          \\
     & $K^{-}/\pi^{-}$  &  5.0 &  4.0 &    5.2 &     &          \\
     & $p/\pi^{+}$  &  7.1 &  6.4 &   7.2 &   7.8  &       \\
     & $\bar{p}/\pi^{-}$  &  7.2 &  6.4 &   6.5 &   7.9  &         \\ \\

\cuau  & \pip  &      10.2  &      10.9  &      10.5  &      9.9  &           \\
     & \pim  &      8.9  &      10.1  &      10.0  &      10.5  &           \\
     & \Kp  &      10.2  &      8.3  &      7.5  &         &           \\
     & \Km  &      10.0  &      7.7  &      8.6  &         &           \\
     & \prot  &      10.3  &      11.5  &      12.4  &      13.1  &      16.5    \\
     & \aprot  &      8.4  &      8.6  &      10.3  &      10.9  &      12.8    \\
     & $K^{+}/\pi^{+}$  &  7.0 &  7.0 &    7.7 &     &          \\
     & $K^{-}/\pi^{-}$  &  7.2 &  7.2 &    8.0 &     &          \\
     & $p/\pi^{+}$  &  8.5 &  8.6 &  10.0 &   10.1  &         \\
     & $\bar{p}/\pi^{-}$  &  9.1 &  8.5 &   8.7 &   8.7  &         \\ \\

\uu & \pip  &      16.8  &      16.5  &      14.7  &      13.7  &           \\
     & \pim  &      6.2  &      8.8  &      9.4  &      16.9  &           \\
     & \Kp  &      10.2  &      8.3  &      8.3  &         &           \\
     & \Km  &      8.8  &      7.6  &      8.5  &         &           \\
     & \prot  &      10.0  &      9.8  &      9.8  &      11.2  &   14.1       \\
     & \aprot  &      10.3  &      10.0  &      9.1  &      10.5  &     15.2   \\
     & $K^{+}/\pi^{+}$  &  6.9 &  7.0 &    7.4 &     &          \\
     & $K^{-}/\pi^{-}$  &  8.0 &  7.0 &    7.8 &     &          \\
     & $p/\pi^{+}$  &  9.0 &  8.5 &   9.0 &   8.6  &          \\
     & $\bar{p}/\pi^{-}$  &  10.0 &  9.6 &  9.0 &   9.2  &        \\
\end{tabular}
 \end{ruledtabular}
\end{table}
\endgroup

\subsection{Invariant spectra}

\begin{figure*}[!tbh]
\includegraphics[width=0.995\linewidth]{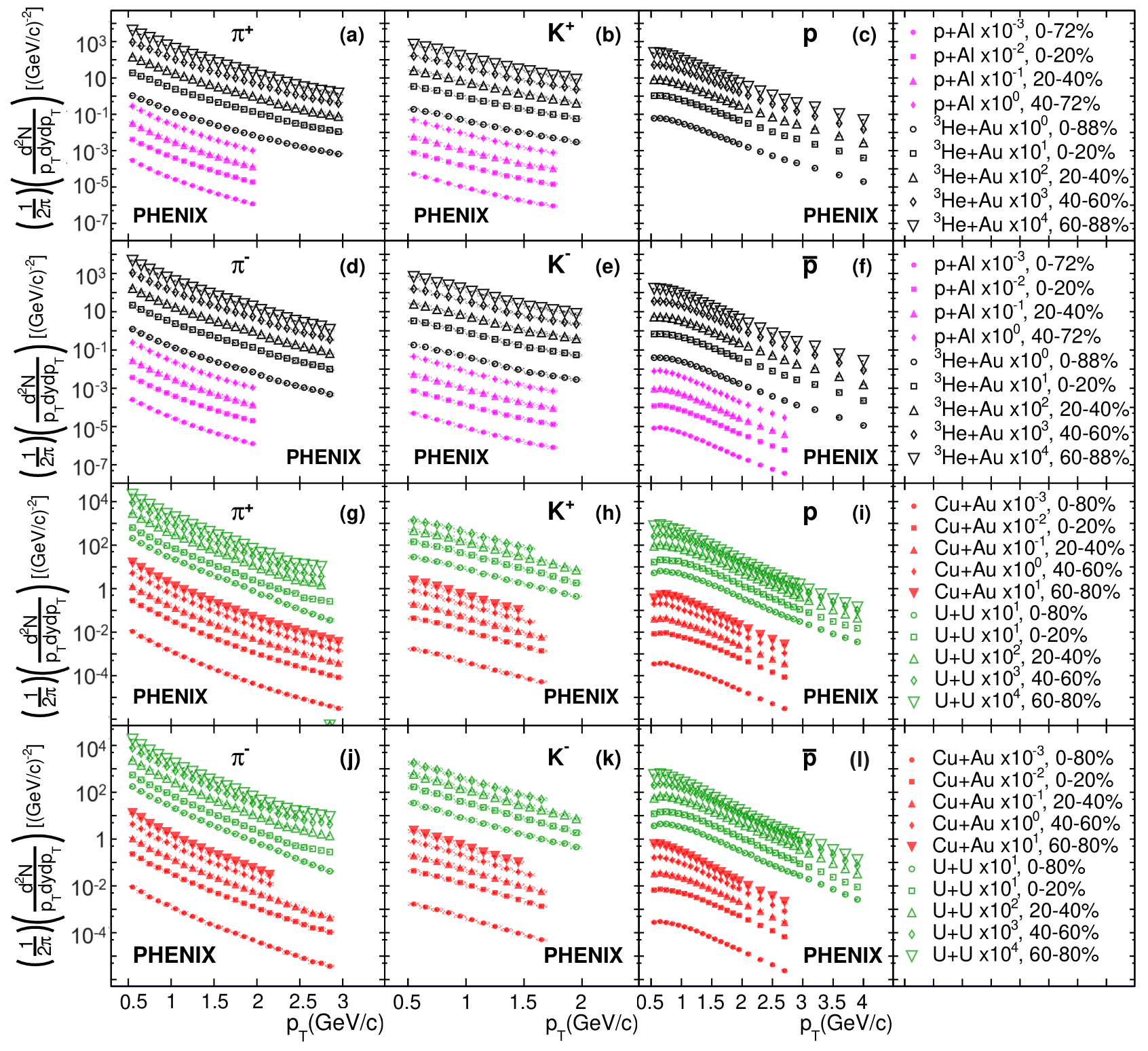}
\caption{The \pip, \pim, \Kp,\Km, \prot and \aprot invariant \pt spectra 
measured in different centralities of \pal, \heau, \cuau collisions at 
\sqsn = 200 GeV and U$+$U collisions at \sqsn = 193 GeV. Invariant \pt 
spectra are multiplied by powers of ten for clarity of presentation.}
\label{fig:spectraPt}
\end{figure*}

The \pipm, \Kpm, \prot, and \aprot invariant transverse momentum spectra 
were measured in \pal, \heau, \cuau collisions at \sqsn = 200 GeV 
and in \uu collisions at \sqsn~=~193 GeV as follows:
\begin{equation}
  \label{eq:spectra}
    \frac{1}{2\pi p_T} \frac{d^2 N}{dp_T dy}=\frac{1}{2\pi p_T}\frac{N_h f_{\rm bias} C_{\rm FD} }{N_{\rm evt} \varepsilon_{\rm rec} \Delta p_T \Delta y},
\end{equation}
where $N_h$ is the raw yield of hadron $h$, $N_{\rm evt}$ is the number of 
nucleus-nucleus collisions, $y$ is rapidity, $C_{\rm FD}$ is the feed-down 
correction and $f_{\rm bias}$ is the bias-factor correction. The 
$C_{\rm FD}=1$ for \pipm and \Kpm; $f_{\rm bias}=1$ for the large 
collision systems. The resulting \pipm, \Kpm, \prot, and \aprot 
invariant \pt spectra are presented in Fig.~\ref{fig:spectraPt}.

The $\pi$, $K$, and $p$ invariant \pt spectra exhibit different shapes. 
To quantify these differences, invariant-transverse-mass 
($m_T=\sqrt{p_T^2 + m_0^2}$) spectra were calculated. The \mt 
invariant spectra of all identified charged hadrons have exponential 
form for \mt $<$ 1.5 GeV and can be approximated by 
Eq.~\ref{eq:spectra_mt}~\cite{PPG026}:

\begin{equation}
  \label{eq:spectra_mt}
    \frac{1}{2\pi m_T} \frac{d^2 N}{dm_T dy}= 
    \frac{A  }{2\pi T (T+m_0)}\exp \left( -\frac{m_T - m_0}{T}\right)
\end{equation}
where $T$ is the inverse-slope parameter and $A$ is a 
normalization factor. Examples of resulting \pip, \Kp and \prot 
invariant-\mt spectra measured in central \cuau collisions are presented 
in Fig.~\ref{fig:spectraMt}. Approximations by Eq.~\ref{eq:spectra_mt} 
are shown with [red] solid lines. Values of inverse-slope parameters 
$T$ calculated in \pal, \heau, Cu$+$Au, and U$+$U collisions are summarized 
in Table~\ref{table:Tinv}.

Figure~\ref{fig:Tinv} shows examples of $T$ parameter vs.~hadron mass 
($m_0$) dependencies for different centralities of Cu$+$Au collisions. 
Ordering of pion, kaon, and proton inverse slope values 
$T_{\pi}<T_{K}<T_p$ can be seen in all centralities.

The difference between $T$ values, calculated for protons in central and 
peripheral U$+$U collisions, is 18\% and decreases from large to 
small collision systems. The $T$ values, calculated for pions in 
different centralities, are nearly of the same values in all collision 
systems. The $T$ values, calculated for kaons, take intermediate values 
between pion and proton $T$-parameter values. Description of hadron \pt 
spectra shape and parameter $T$ dependence on hadron mass, collision 
system size and centrality can be given by thermal 
models~\cite{PPG026,Thermal1,Thermal2}. Thermal models consider a 
thermalized system with a well-defined temperature and a common 
transverse velocity field. Under the assumption of complete decoupling 
between the thermal and collective motion of the particles, the hadron 
kinetic energy ($\left< E_{\rm kinetic} \right>$) is equal to the linear 
sum of the energy of the thermal motion 
($\left< E_{\rm thermal} \right>$) and the energy of the collective 
motion ($\left< E_{\rm collective} \right>$), 
$\left< E_{\rm kinetic} \right> = \left< E_{\rm thermal} \right> + 
\left< E_{\rm collective} \right>$.  For that reason, the inverse-slope 
parameter $T$ exhibits mass dependence given by 


\begin{equation} 
\label{eq:InvSlope} 
T = T_0 +m \left< u_t\right>^2, 
\end{equation} 
where $T_0$ can be interpreted as a freeze-out temperature and \ut as 
the average collective velocity for all particle species.

\section{RESULTS}

\begin{figure}[!tbh]
\begin{minipage}{0.995\linewidth}
\includegraphics[width=1.0\linewidth]{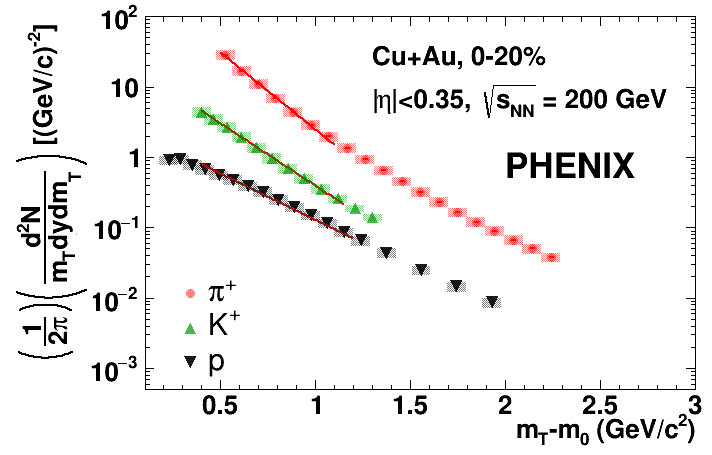}
\caption{An example of \pip, \Kp, \prot invariant \mt spectra measured 
in central \cuau collisions at \sqsn = 200 GeV. Approximations with 
Eq.~\ref{eq:spectra_mt} are presented with red solid lines.}
\label{fig:spectraMt}
\end{minipage}
\begin{minipage}{0.995\linewidth}
\vspace{0.5cm}
\includegraphics[width=1.0\linewidth]{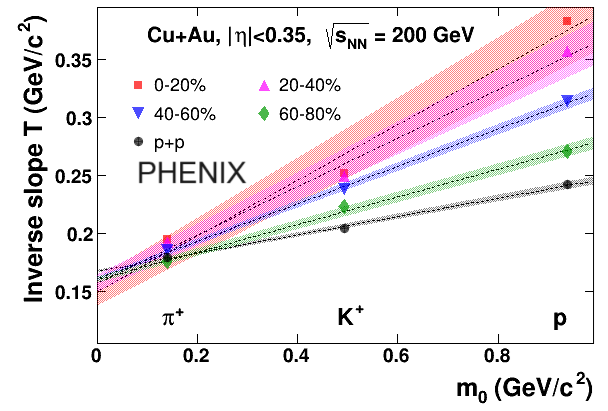}
\caption{Mass and centrality dependence of inverse slope parameters $T$ 
for \pip, \Kp and \prot in Cu$+$Au collisions at \sqsn~=~200 GeV. The 
dotted lines represent a linear fits of the $T(m_0)$ values from each 
centrality bin by Eq.~\ref{eq:InvSlope}. The $T$ values measured for 
\pip, \Kp, and \prot in \pp collisions are shown for comparison.}
\label{fig:Tinv}
\end{minipage}
\end{figure}

\begin{figure}[!tbh]
\begin{minipage}{0.995\linewidth}
\includegraphics[width=1.0\linewidth]{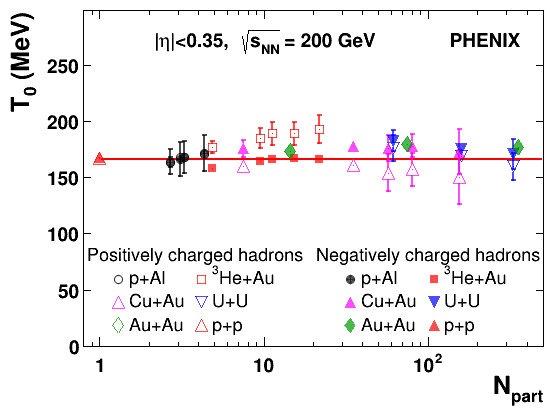}
\caption{Freeze-out temperature (\To) as a function of \Npart obtained 
for positively and negatively charged hadrons at different centralities 
of \pal, \heau, Cu$+$Au, and U$+$U collisions. The \To values measured 
in \pp collisions are shown for comparison.}
\label{fig:T0Npart}
\end{minipage}
\begin{minipage}{0.995\linewidth}
\vspace{0.5cm}
\includegraphics[width=1.0\linewidth]{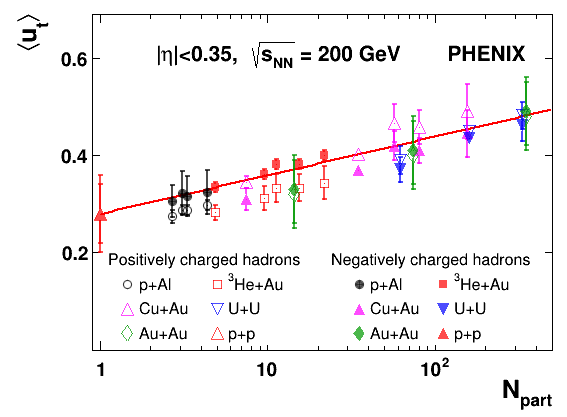}
\caption{Collective velocities ($\left< u_t \right>$) as a function of 
\Npart obtained for positively and negatively charged hadrons in 
different centralities of \pal, \heau, Cu$+$Au, and U$+$U collisions. 
The $\left< u_t \right>$ values measured in \pp collisions are shown for 
comparison.}
\label{fig:utNpart}
\end{minipage}
\end{figure}

\begin{table*}[!tbh]
\begin{minipage}{0.995\linewidth}
\caption{Identified charged hadron inverse slope parameters $T$ 
(MeV/$c^2$) calculated in \pal, \heau, Cu$+$Au, U$+$U collision 
systems.}
\begin{ruledtabular} \begin{tabular}{ccccccccc}
System & \Npart  &  \pip & \pim & \Kp & \Km & \prot & \aprot     \\ \hline
\pal 
&3.1 $\pm$ 0.1 &  178.88 $\pm$ 0.35  &  187.36 $\pm$ 0.18  &  210.69 $\pm$ 0.77  &  206.61 $\pm$ 0.78  &     &  269.51 $\pm$ 1.25    \\
&4.4 $\pm$ 0.3 &  183.83 $\pm$ 0.28  &  192.56 $\pm$ 0.29  &  216.10 $\pm$ 1.22  &  211.25 $\pm$ 1.21  &  &  275.14 $\pm$ 1.16    \\
&3.3 $\pm$ 0.1 &  178.20 $\pm$ 0.32  &  186.96 $\pm$ 0.33  &  210.05 $\pm$ 1.43  &  206.25 $\pm$ 1.46  &   &  266.62 $\pm$ 2.28    \\
&1.6 $\pm$ 0.2  &  173.88 $\pm$ 0.46  &  181.99 $\pm$ 0.26  &  204.74 $\pm$ 1.16  &  201.47 $\pm$ 1.18  &   &  254.20 $\pm$ 1.82    \\
    \\
\heau   
&11.3 $\pm$ 0.5  &  208.70 $\pm$ 0.07  &  187.25 $\pm$ 0.06  &  235.84 $\pm$ 0.24  &  236.98 $\pm$ 0.25  &  295.74 $\pm$ 0.20 &  303.59 $\pm$ 0.24    \\
&21.1 $\pm$ 1.3  &  214.33 $\pm$ 0.11  &  191.58 $\pm$ 0.09  &  242.57 $\pm$ 0.38  &  242.91 $\pm$ 0.39  &  309.27 $\pm$ 0.33  &  317.40 $\pm$ 0.76    \\
&15.4 $\pm$ 0.9  &  209.84 $\pm$ 0.13  &  188.33 $\pm$ 0.10  &  237.27 $\pm$ 0.44  &  238.49 $\pm$ 0.46  &  296.44 $\pm$ 0.37  &295.10 $\pm$ 0.18    \\
&9.5 $\pm$ 0.6  &  202.71 $\pm$ 0.15  &  182.66 $\pm$ 0.13  &  227.37 $\pm$ 0.53  &  229.34 $\pm$ 0.55  &  280.01 $\pm$ 0.44  &  287.12 $\pm$ 0.54    \\
&4.8 $\pm$ 0.3 &  191.01 $\pm$ 0.18  &  172.40 $\pm$ 0.15  &  213.44 $\pm$ 0.66  &  215.99 $\pm$ 0.70  &  254.22 $\pm$ 0.55  &  261.83 $\pm$ 0.66    \\
    \\
\cuau
&70.4 $\pm$ 3.0  &  191.72 $\pm$ 0.01  &  205.41 $\pm$ 0.01  &  249.33 $\pm$ 0.03  &  253.80 $\pm$ 0.03  &  363.65 $\pm$ 0.09  &  344.75 $\pm$ 0.08    \\
&154.8 $\pm$ 4.1  &  194.29 $\pm$ 0.12  &  208.57 $\pm$ 0.01  &  251.43 $\pm$ 0.04  &  255.85 $\pm$ 0.09  &  383.85 $\pm$ 0.13  &  365.18 $\pm$ 0.13    \\
&80.4 $\pm$ 3.3  &  192.07 $\pm$ 0.02  &  205.67 $\pm$ 0.02  &  249.14 $\pm$ 0.06  &  253.34 $\pm$ 0.05  &  357.54 $\pm$ 0.16  &  338.92 $\pm$ 0.15    \\
&34.9 $\pm$ 1.8  &  185.32 $\pm$ 0.03  &  197.72 $\pm$ 0.03  &  238.05 $\pm$ 0.09  &  243.35 $\pm$ 0.09  &  313.90 $\pm$ 0.20  &  306.48 $\pm$ 0.20    \\
&11.5 $\pm$ 1.8  &  175.24 $\pm$ 0.05  &  186.06 $\pm$ 0.05  &  222.76 $\pm$ 0.18  &  228.90 $\pm$ 0.18  &  270.86 $\pm$ 0.28  &  263.38 $\pm$ 0.29    \\
\\
\uu 
& 330 $\pm$ 6 &  198.18 $\pm$ 0.02  &  205.84 $\pm$ 0.21  &  266.49 $\pm$ 0.08  &  265.80 $\pm$ 0.08  &  382.54 $\pm$ 0.19  &  374.48 $\pm$ 0.21    \\
& 259 $\pm$ 7 &  197.28 $\pm$ 0.04  &  202.39 $\pm$ 0.03  &  269.42 $\pm$ 0.12  &  266.36 $\pm$ 0.11  &  358.24 $\pm$ 0.23  &  353.94 $\pm$ 0.26    \\
& 65 $\pm$ 6  &  192.42 $\pm$ 0.06  &  197.74 $\pm$ 0.06  &  259.69 $\pm$ 0.22  &  257.80 $\pm$ 0.20  &  314.92 $\pm$ 0.32  &  308.57 $\pm$ 0.35    \\
\end{tabular} \end{ruledtabular}
\label{table:Tinv}
\end{minipage}
\begin{minipage}{0.995\linewidth}
\vspace{0.5cm}
\caption{Freeze-out temperatures and averaged collective velocities 
calculated for positively charged ($T_{0}^{+}$, 
$\left<u_{t}\right>^{+}$) and negatively charged ($T_{0}^{-}$, 
$\left<u_{t}\right>^{-}$) hadrons at different centralities of \pal, 
\heau, Cu$+$Au, and U$+$U collisions}.
\begin{ruledtabular} \begin{tabular}{cccccc}
System &\Npart     & $T_{0}^{+}$ & $T_{0}^{-}$  & $\left<u_{t}\right>^{+}$ & $\left<u_{t}\right>^{-}$   \\ \hline
\pal
& 3.1$\pm$0.1&  167.99 $\pm$ 2.17  &  166.49 $\pm$ 15.29  &  0.28 $\pm$ 0.01  &  0.32 $\pm$ 0.04    \\
&4.4$\pm$0.3  &  171.84 $\pm$ 0.92  &  171.33 $\pm$ 15.96  &  0.29 $\pm$ 0.01  &  0.32 $\pm$ 0.04  \\
&3.3$\pm$0.1 &  168.01 $\pm$ 2.15  &  167.88 $\pm$ 14.30  &  0.28 $\pm$ 0.01  &  0.31 $\pm$ 0.04    \\
&1.6$\pm$0.2 &  164.10 $\pm$ 4.39  &  163.80 $\pm$ 11.20  &  0.27 $\pm$ 0.01  &  0.30 $\pm$ 0.03    \\
\\
\heau 
&11.3$\pm$0.5 &  189.16 $\pm$ 10.22  &  166.27 $\pm$ 1.70  &  0.33 $\pm$ 0.03  &  0.38 $\pm$ 0.01    \\
&21.1$\pm$1.3 &  193.84 $\pm$ 12.42  &  166.44 $\pm$ 3.23  &  0.34 $\pm$ 0.03  &  0.39 $\pm$ 0.01    \\
&15.4$\pm$0.9 &  188.99 $\pm$ 9.82  &  167.59 $\pm$ 1.24  &  0.33 $\pm$ 0.02  &  0.38 $\pm$ 0.01    \\
&9.5$\pm$0.6 &  185.05 $\pm$ 8.57  &  164.60 $\pm$ 0.27  &  0.31 $\pm$ 0.02  &  0.36 $\pm$ 0.01    \\
&4.8$\pm$0.3  &  177.02 $\pm$ 4.99  &  158.33 $\pm$ 3.45  &  0.28 $\pm$ 0.01  &  0.33 $\pm$ 0.01    \\
\\
Cu$+$Au 
&70.4$\pm$3.0  &  154.06 $\pm$ 16.63  &  176.03 $\pm$ 11.95  &  0.46 $\pm$ 0.03  &  0.41 $\pm$ 0.02    \\
&154.8$\pm$4.1  &  149.99 $\pm$ 23.98  &  172.82 $\pm$ 19.75  &  0.48 $\pm$ 0.05  &  0.44 $\pm$ 0.04    \\
&80.4$\pm$3.3   &  157.02 $\pm$ 14.55  &  178.11 $\pm$ 10.20  &  0.45 $\pm$ 0.03  &  0.40 $\pm$ 0.02    \\
&34.9$\pm$2.9  &  161.02 $\pm$ 3.86  &  177.79 $\pm$ 2.35  &  0.40 $\pm$ 0.01  &  0.36 $\pm$ 0.01    \\
&11.5$\pm$1.8  &  160.04 $\pm$ 4.50  &  175.86 $\pm$ 7.56  &  0.34 $\pm$ 0.01  &  0.30 $\pm$ 0.02    \\
\\
U$+$U 
&330$\pm$6  &  160.03 $\pm$ 12.02  &  170.80 $\pm$ 13.24  &  0.48 $\pm$ 0.02  &  0.46 $\pm$ 0.03    \\
&259$\pm$7  &  169.04 $\pm$ 0.60  &  174.76 $\pm$ 2.93  &  0.44 $\pm$ 0.01  &  0.43 $\pm$ 0.01    \\
&65$\pm$6  &  176.03 $\pm$ 11.45  &  182.61 $\pm$ 9.64  &  0.39 $\pm$ 0.02  &  0.37 $\pm$ 0.02    \\
\end{tabular} \end{ruledtabular}
\label{table:T0ut}
\end{minipage}
\end{table*}

The dotted lines on Fig.~\ref{fig:Tinv} represent linear fits of 
the $T(m_0)$ values from each centrality bin by 
Eq.~\ref{eq:InvSlope}. The fit parameters for positively charged 
($T_{0}^{+}$, $\left<u_{t}\right>^{+}$) and negatively charged 
($T_{0}^{-}$, $\left<u_{t}\right>^{-}$) hadrons calculated in 
\pal, \heau, Cu$+$Au, and U$+$U collision systems are shown in 
the Table~\ref{table:T0ut} and presented in 
Figs.~\ref{fig:T0Npart}~and~\ref{fig:utNpart} as a function of 
\Npart values. Presented error values for $T_{0}^{\pm}$ and 
$\left<u_{t}\right>^{\pm}$ parameters are due to the fit 
uncertainties. The $T_0$ values calculated in collisions with 
different geometries and centralities were found to be coincident 
within uncertainties, indicating that the freeze-out temperature 
is approximately independent of \Npart values. The averaged $T_0$ 
value was found to be $166.1 \pm 2.2$ MeV, which is shown as the 
[red] solid line in Fig.~\ref{fig:T0Npart}.  In contrast, the 
strength of the average transverse flow (\ut values) increases 
from small to large \Npart values, supporting the hydrodynamic 
picture. The $\left< u_t \right>(\left< N_{\rm part} \right>)$ 
values were approximated with the fit function $p_1 \cdot log(p_2 
\cdot \left< N_{\rm part} \right>)$, where 
$p_1 = 0.0345{\pm}0.0003$ and $p_2=3196{\pm}342$. These fit parameters 
could provide insight to theoretical models for radially expanding 
thermalized systems.


\begin{figure*}[!tbh]
\begin{minipage}{0.995\linewidth}
\includegraphics[width=0.99\linewidth]{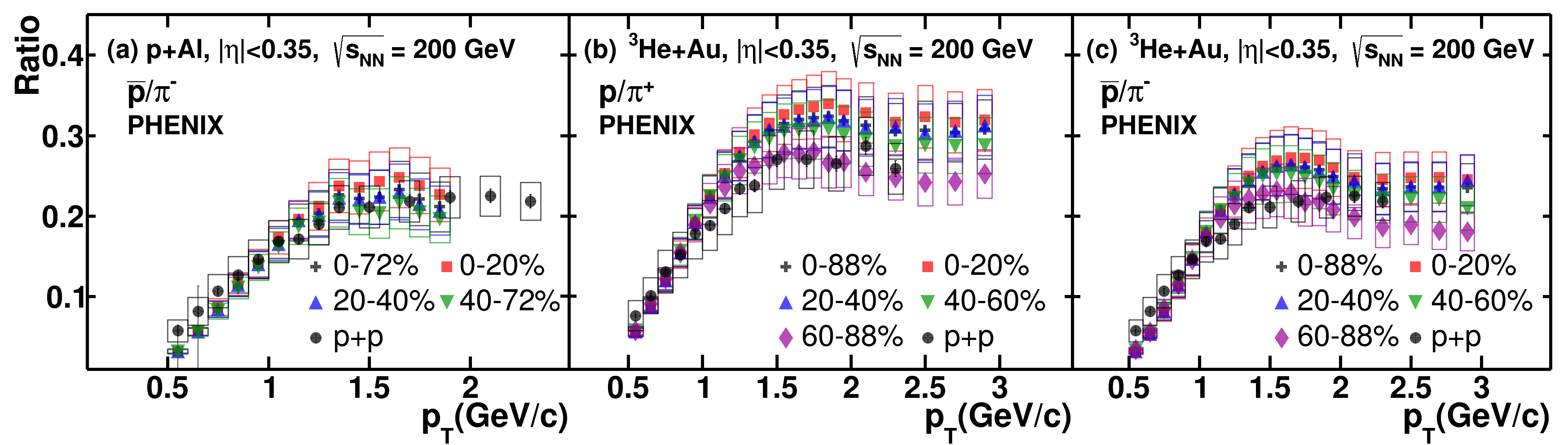}
\caption{The ratios of \prot/\pip as a function of \pt measured in 
different centralities of \heau collisions and \aprot/\pim ratios 
measured in different centralities of \pal and \heau collisions.  Data 
points, measured in \pp collisions~\cite{CH_pp}, are shown as a 
comparison.}
\label{fig:p2pi_small}
\end{minipage}
\begin{minipage}{0.995\linewidth}
\vspace{0.4cm}
\includegraphics[width=0.7\linewidth]{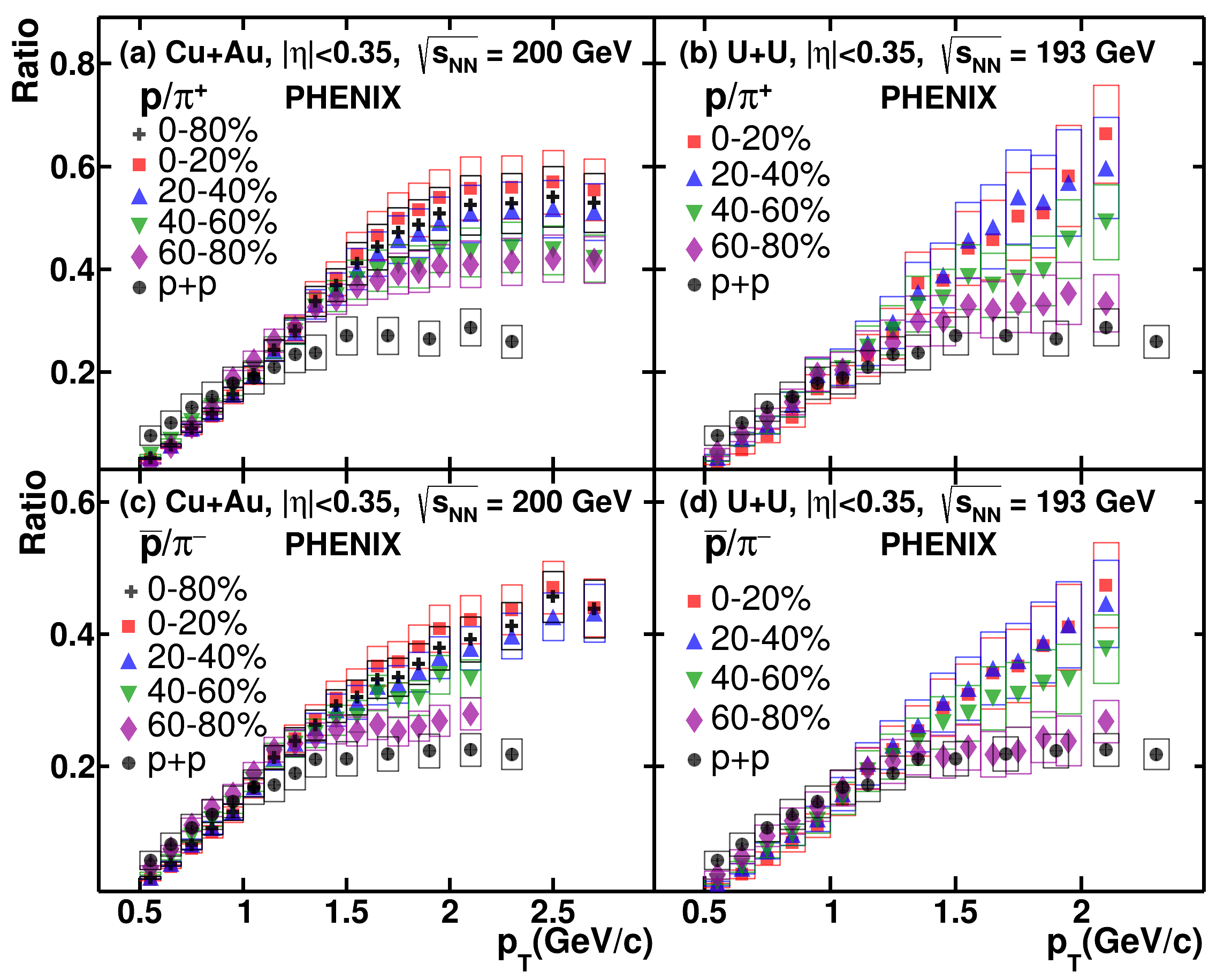}
\caption{The ratios of \prot/\pip and \aprot/\pim as a function of \pt 
measured in different centralities of Cu$+$Au, and U$+$U collisions. 
Data points measured in \pp collisions~\cite{CH_pp} are shown for 
comparison. }
\label{fig:p2pi_large}
\end{minipage}
\end{figure*}

\begin{figure*}[!tbh]
\begin{minipage}{0.995\linewidth}
\includegraphics[width=0.7\linewidth]{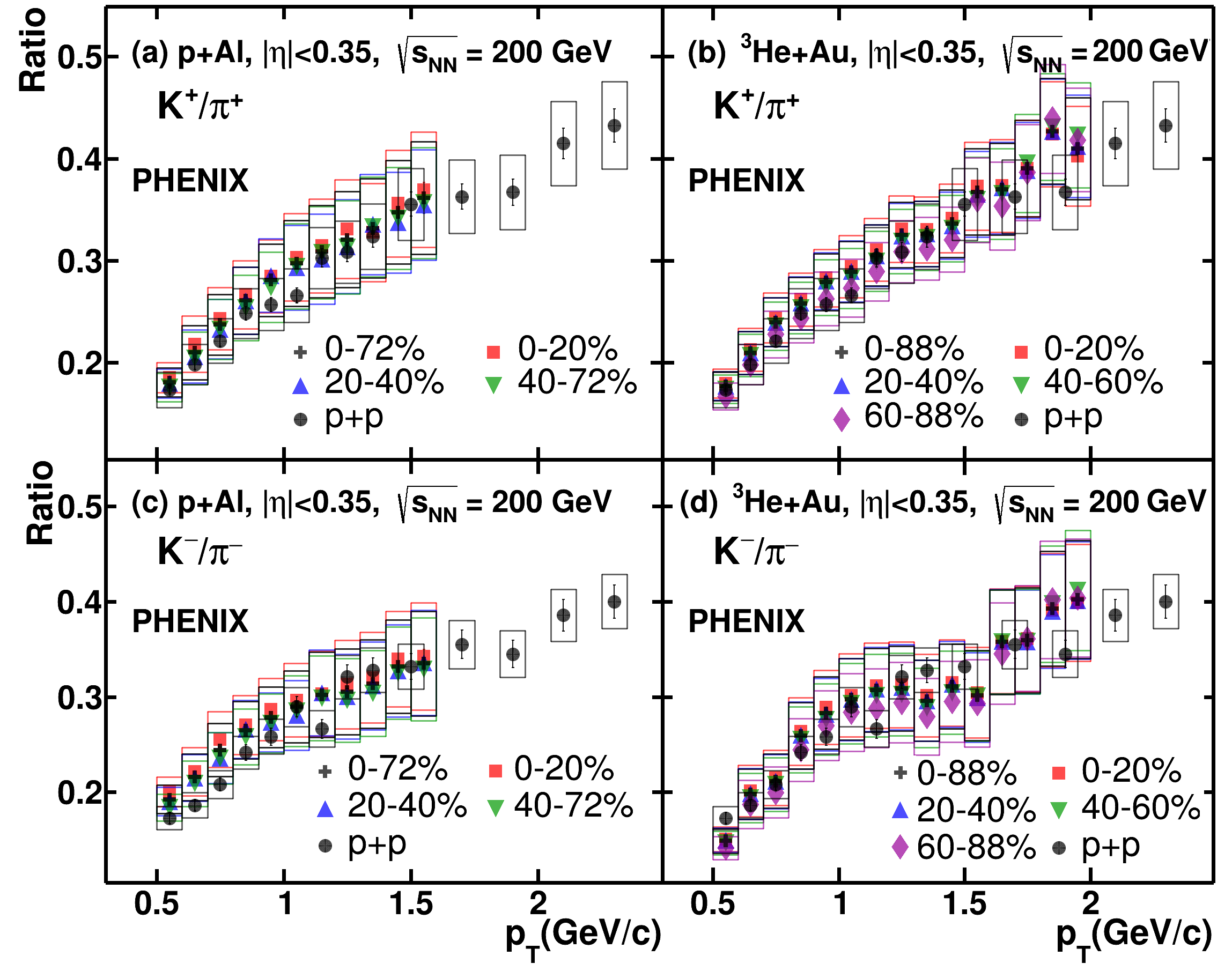}
\caption{The ratios of \Kp/\pip and \Km/\pim as a function of \pt 
measured in different centralities of \pal and \heau collisions.  Data 
points measured in \pp collisions~\cite{CH_pp} are shown for 
comparison.}
\label{fig:K2pi_small}
\end{minipage}
\begin{minipage}{0.995\linewidth}
\vspace{1.0cm}
\includegraphics[width=0.7\linewidth]{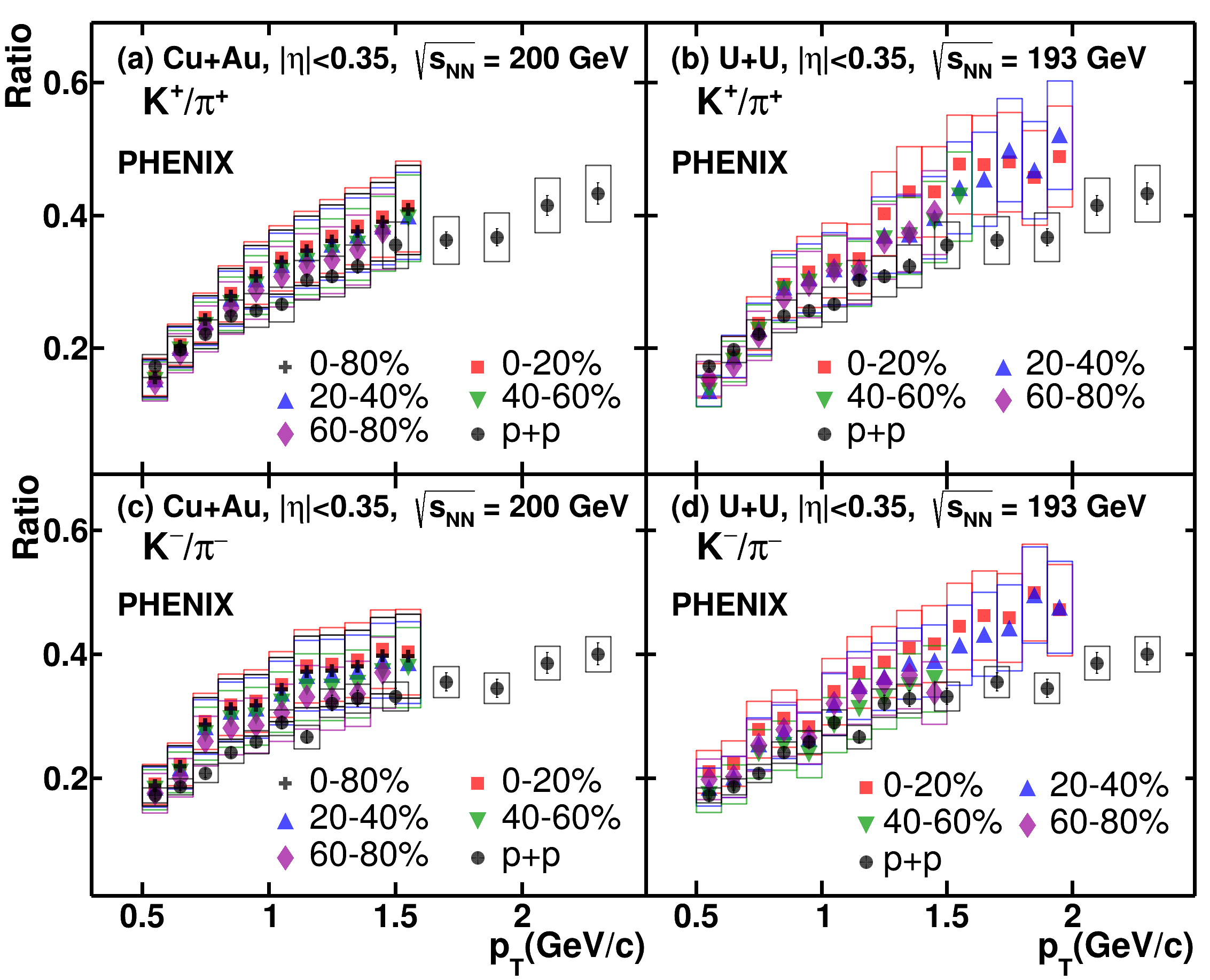} 
\caption{The ratios of \Kp/\pip and \Km/\pim as a function of \pt 
measured in different centralities of Cu$+$Au, and U$+$U collisions. 
Data points measured in \pp collisions~\cite{CH_pp} are shown for 
comparison.}
\label{fig:K2pi_large}
\end{minipage}
\end{figure*}

\begin{figure*}[!tbh]
\begin{minipage}{0.995\linewidth}
\includegraphics[width=0.99\linewidth]{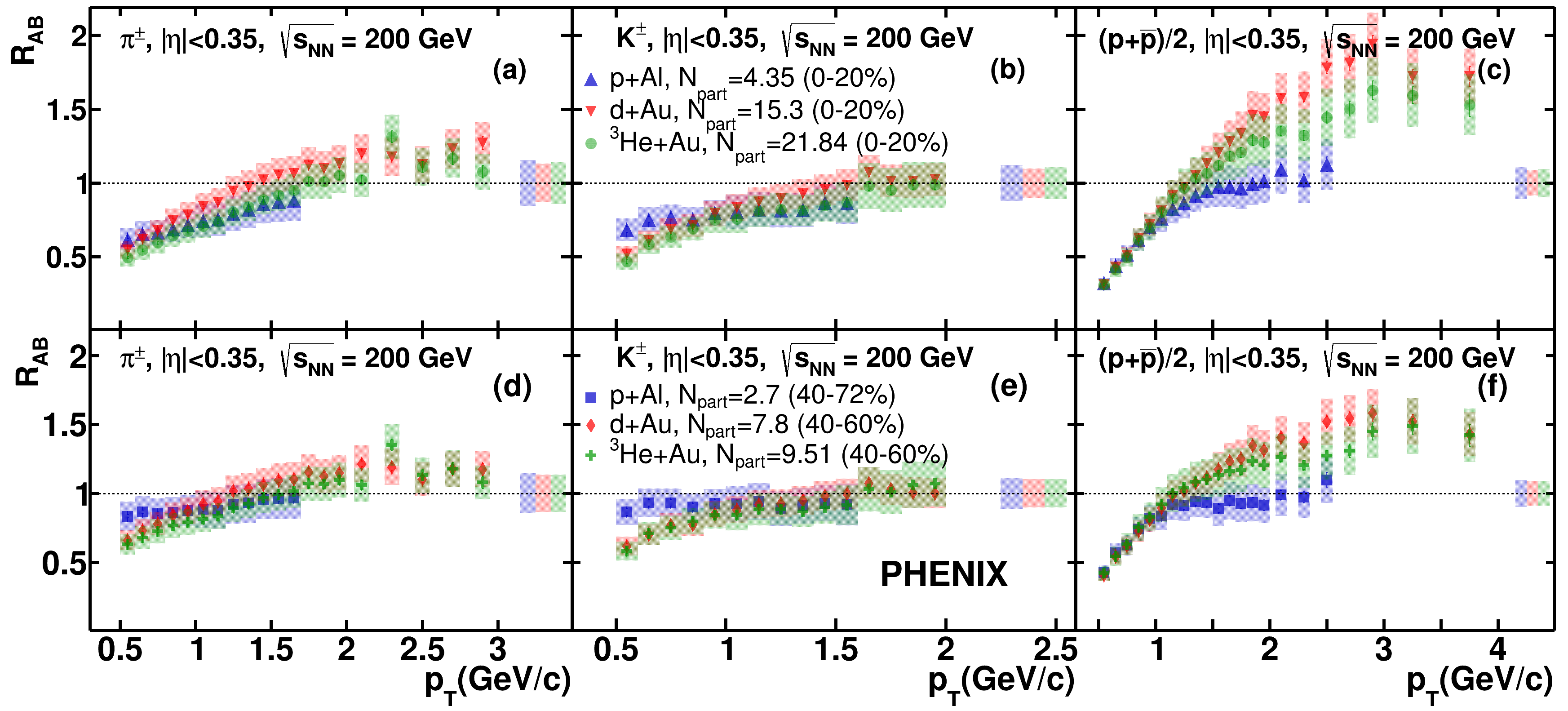}
\caption{Identified charged hadron nuclear-modification factors as a 
function of \pt measured in central and peripheral \pal, \dau and \heau 
collisions. The dashed lines are drawn as a visual aid at the value of 
\rab = 1 indicating absence of nuclear modification.}
\label{fig:RAB_small}
\end{minipage}
\begin{minipage}{0.995\linewidth}
\vspace{0.4cm}
\includegraphics[width=0.99\linewidth]{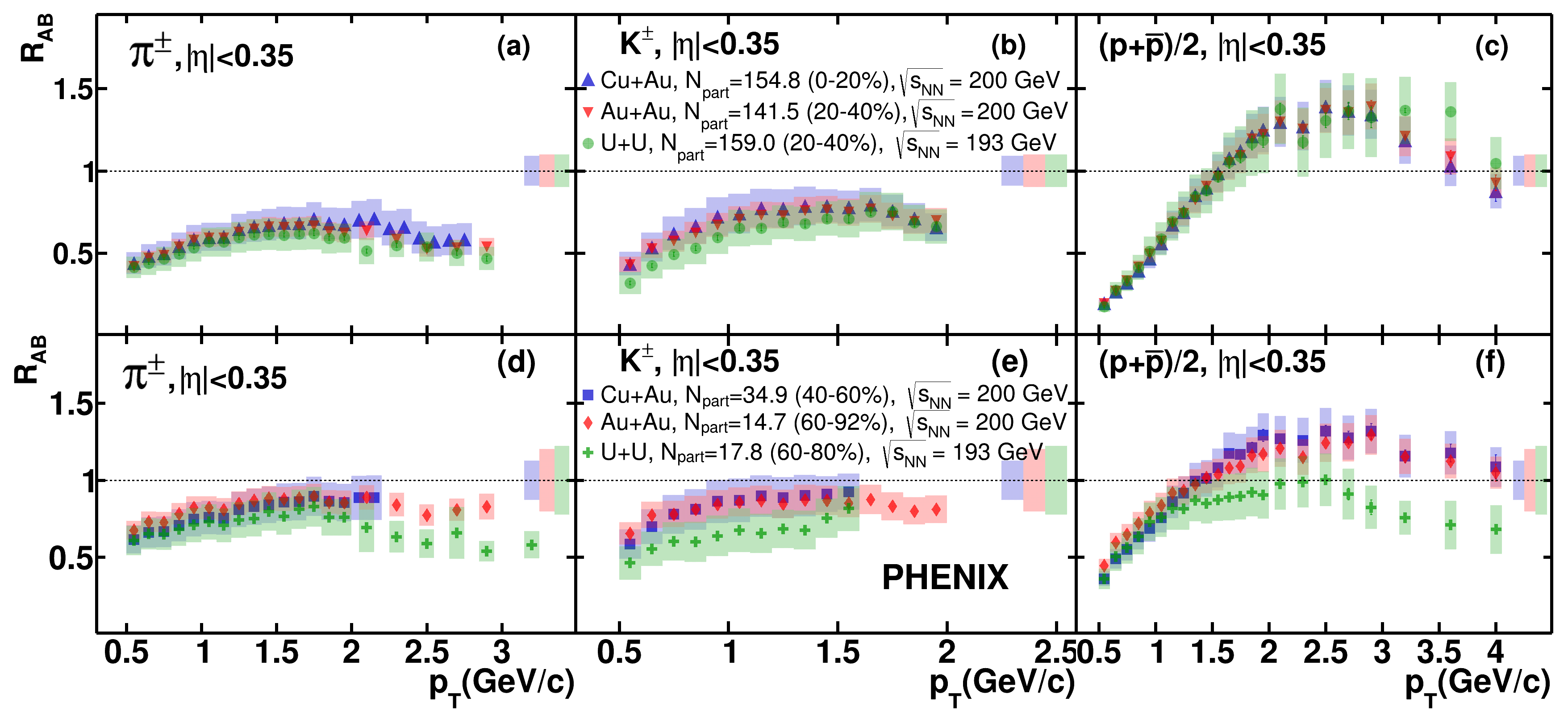}
\caption{Identified charged hadron nuclear-modification factors as a 
function of \pt measured in central and peripheral Cu$+$Au, Au$+$Au and 
U$+$U collisions. The dashed lines are drawn as a visual aid at the 
value of \rab = 1 indicating absence of nuclear modification.}
\label{fig:RAB_large}
\end{minipage}
\end{figure*}

\begin{figure*}[tbh]
\begin{minipage}{0.995\linewidth}
\includegraphics[width=0.65\linewidth]{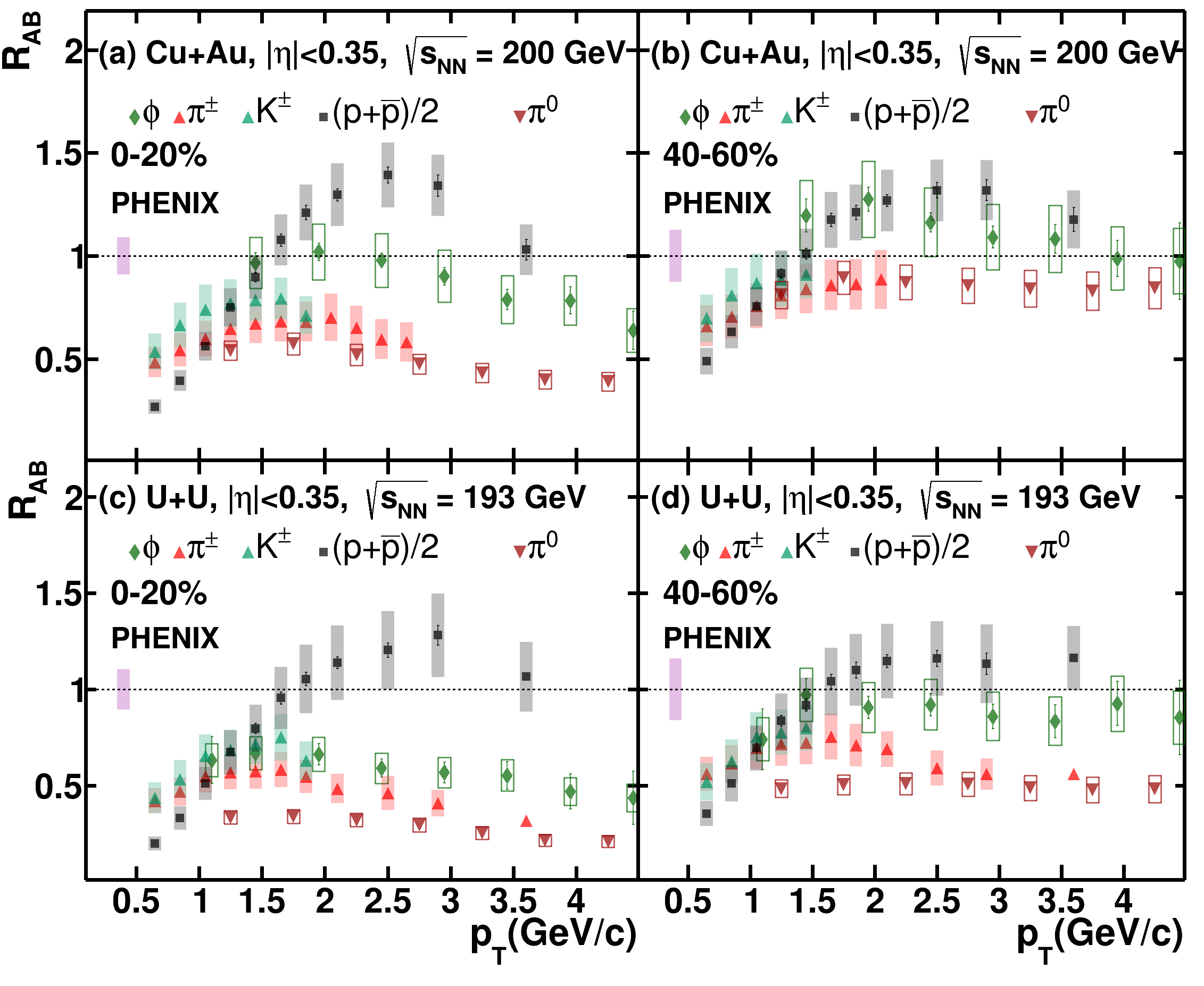}
\caption{Light hadron ($\phi$~\cite{phi_LargeSysts}, \pipm, \Kpm, \prots 
and $\pi^0$~\cite{pi0Eta_CuAu,pi0Eta_UU}) \rab values vs.~\pt measured 
in central and peripheral Cu$+$Au, and U$+$U collisions. The dashed 
lines are drawn as a visual aid at the value of \rab = 1 indicating 
absence of nuclear modification. }
\label{fig:RabMesons_large}
\end{minipage}
\begin{minipage}{0.995\linewidth}
\vspace{0.5cm}
\includegraphics[width=0.65\linewidth]{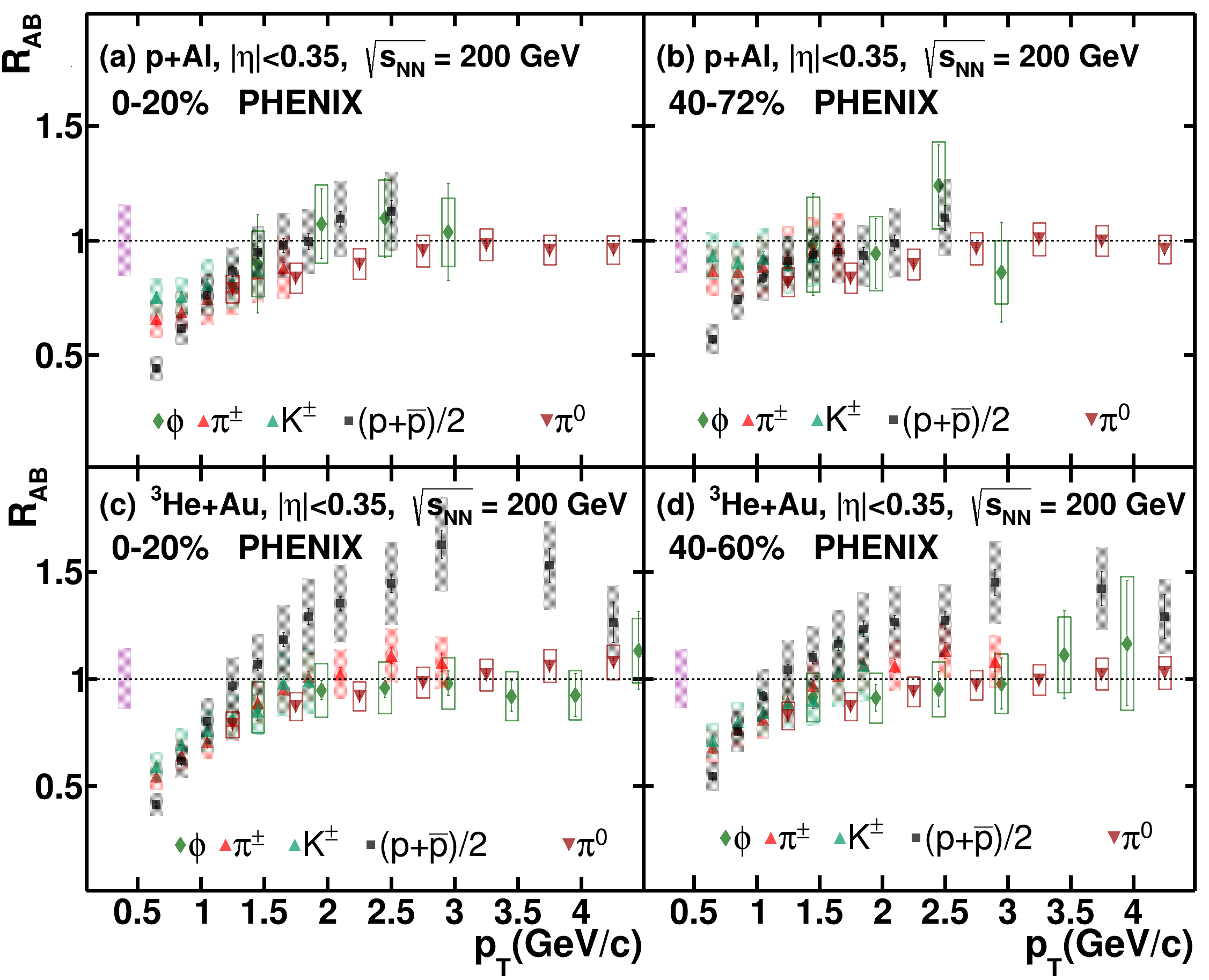}
\caption{Comparison of $\phi$~\cite{phi_SmallSysts,phi_dAu}, \pipm, 
\Kpm, \prots and $\pi^0$~\cite{pi0_smallSysts} \rab values vs.~\pt 
measured in central and peripheral $p$$+A$l and \heau collisions. The 
dashed lines are drawn as a visual aid at the value of \rab = 1 
indicating absence of nuclear modification.}
\label{fig:RabMesons_small}
\end{minipage}
\end{figure*}

\subsection{Particle ratios}

Enhancement of baryon production in nucleus-nucleus collisions is 
considered to be one of the signatures of QGP 
formation~\cite{PPG026,CHspectra_130GeV,ppg146}. To 
investigate differences in baryon and meson production mechanisms, the 
ratios of $p/\pi$ and $K/\pi$ have been calculated. 
Figures~\ref{fig:p2pi_small}~and~\ref{fig:p2pi_large} present 
comparisons of \prot/\pip and \aprot/\pim ratios in different 
centralities of large (\cuau, \uu) and small (\pp, \pal, \heau) 
collision systems. In central collisions of large systems $p$/$\pi$ 
ratios reach the values of $\approx\,0.6$, but in peripheral collisions the 
values of $p/\pi$ ratios are smaller than 0.4 in the whole \pt range. 
Behavior of $p/\pi$ ratios observed in \cuau and \uu collision systems 
can be qualitatively described using recombination 
models~\cite{Recombination1,Recombination2}.

In small collision systems (\pal, \heau), the values of $p$/$\pi$ ratios 
are similar to those measured in \pp collisions~\cite{CH_pp}. In \heau 
collisions a modest centrality dependence can be seen, similar to that 
observed in \dau collisions~\cite{PPG026,ppg146}. The modest centrality 
dependence can be understood in terms of the small range of \Npart values 
relative the large range of \Npart values in large collision systems. 
The values of $p/\pi$ ratios measured in all centrality classes of \pal 
collisions and in \pp collisions are consistent within uncertainties. 
For all measurements type A uncertainties sum quadratically to type B 
uncertainties and are shown as a rectangles around the experimental 
points in Figs.~\ref{fig:K2pi_small}--\ref{fig:RabMesons_small}.

The ratios of \Kp/\pip and \Km/\pim are shown in 
Figs.~\ref{fig:K2pi_small}~and~\ref{fig:K2pi_large}. The values of 
$K/\pi$ ratios show a modest centrality dependence, which is 
insignificant within systematic uncertainties. The centrality dependence 
of $K/\pi$ ratios in \dau and \auau collisions was attributed to 
a strangeness-enhancement effect~\cite{ppg146}.

\subsection{Nuclear-modification factors}

To quantify differences of hadron production in relativistic 
nucleus-nucleus collisions (A+B) and in \pp collisions, nuclear 
modification factors (\rab) were calculated as follows:
\begin{equation}
  \label{eq:rab}
    R_{AB}=\frac{1}{\left< N_{\rm coll} \right>}\frac{d^2 N_{A+B}/dp_T dy}{d^2 N_{p+p}/dp_T dy},
\end{equation}
where $d^2 N_{A+B}/dp_T dy$ and $d^2 N_{p+p}/dp_T dy$ are the invariant 
spectra measured in A+B and \pp collisions, respectively.

Figures~\ref{fig:RAB_small}~and~\ref{fig:RAB_large} present comparisons 
of identified charged-hadron \rab values as a function of \pt in central 
and peripheral \pal, \heau, Cu$+$Au, and U$+$U collisions. The \rab values 
are found to be in agreement in collisions with different geometries, 
but with the same \Npart values, indicating that identified 
charged-hadron production depends only on system size and not geometry.

The following features of identified charged-hadron production in \pal 
collisions have been found: (i) the slope of $R_{AB}(p_T)$ in \pal 
collisions is flatter than it is in \heau and \dau collisions and (ii) 
proton \rab values in \pal collisions at the intermediate \pt range 
(1.0~GeV/$c$~$<~p_T~<$~2.5 GeV/$c$) are equal to unity, while in \heau and 
\dau collisions proton \rab is above unity. Differences in identified 
charged-hadron production between \pal and $d/$\heau might be caused by 
the size of the \pal system being insufficient to observe an increase in 
proton production.

Comparisons of identified charged-hadron \rab values with 
neutral-meson \rab 
values~\cite{pi0_smallSysts,pi0Eta_CuAu,pi0Eta_UU} in small and 
large collision systems are presented in 
Figs.~\ref{fig:RabMesons_large}~and~\ref{fig:RabMesons_small}. In 
large collision systems and in the $^{3}$He+Au collision system, 
proton \rab values are enhanced over all meson \rab values. The 
mass of the $\phi$-meson $m_{\phi}$=1019 MeV/$c^2$ is similar to 
the proton mass $m_{p}$=938 MeV/$c^2$, therefore the enhancement 
of proton $R_{AB}$ values over $\phi$-meson $R_{AB}$ values 
suggests differences in baryon versus meson production instead of 
a simple mass dependence. In \pal collisions proton $R_{AB}$ 
values and $R_{AB}$ values of all measured mesons are in 
agreement within uncertainties, which shows zero enhancement in 
proton to $\phi$-meson production.

\begin{figure*}[!tbh]
\includegraphics[width=0.695\linewidth]{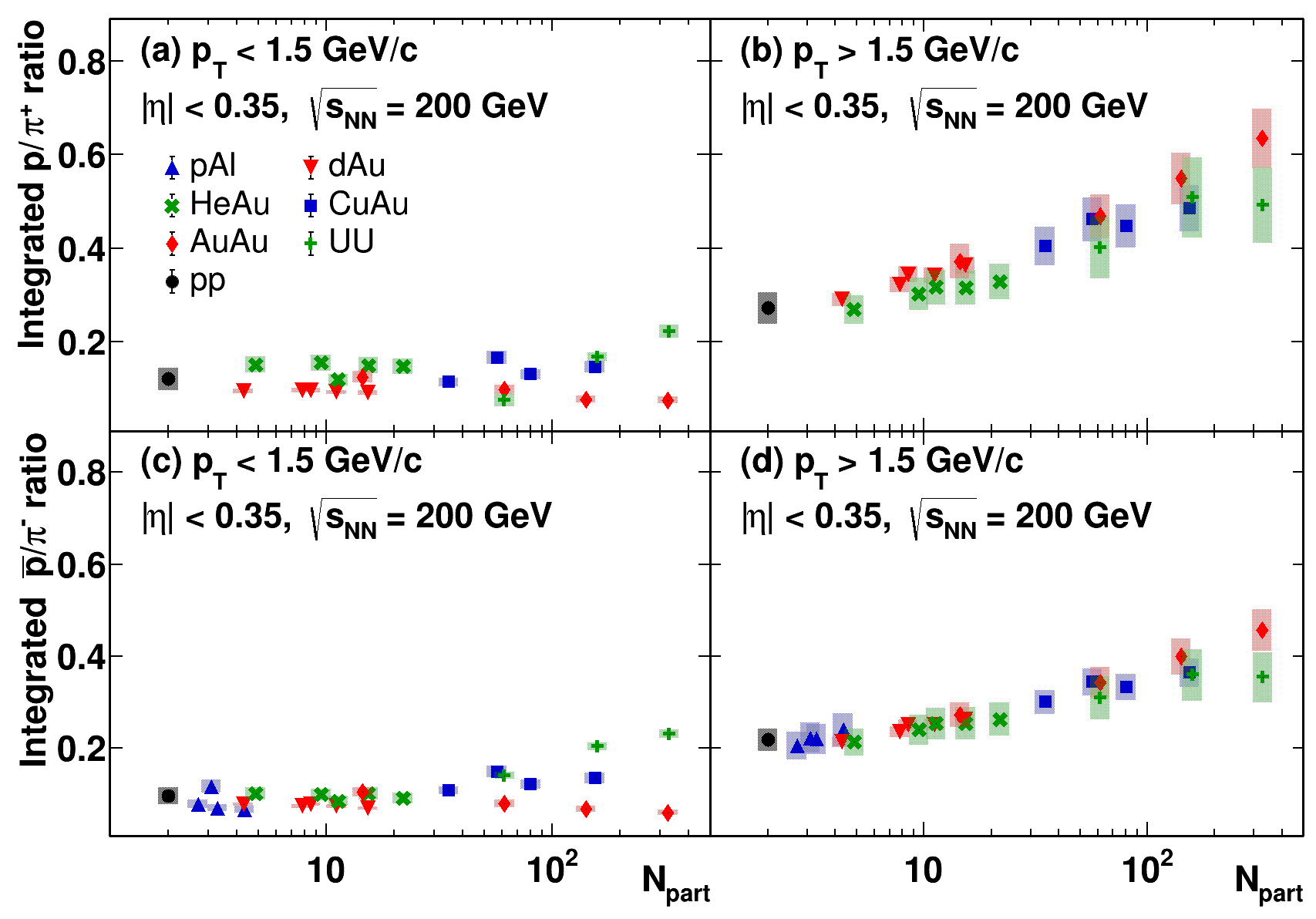}
\caption{The \pip, \pim, \Kp,\Km, \prot and \aprot invariant \pt spectra 
measured in different centralities of \pal, \heau, \cuau collisions at 
\sqsn = 200 GeV and U$+$U collisions at \sqsn = 193 GeV. Invariant \pt 
spectra are multiplied by powers of ten for clarity of presentation.}
\label{fig:IntegratedRatios}
\end{figure*}


The $p/\pi^{+}$ and $\bar{p}/\pi^{-}$ ratios integrated in the 
low-\pt (high-\pt) regions, $p_T < 1.5$ GeV/$c$ ($p_T > 1.5$ 
GeV/$c$) are plotted in Fig.~\ref{fig:IntegratedRatios} as a 
function of $\left< N_{\rm part} \right>$.  The values of ratios, 
integrated in the low-$p_T$ region 
[Fig.~\ref{fig:IntegratedRatios}(a)~and~\ref{fig:IntegratedRatios}(c)], 
were found to be approximately independent of 
$\left< N_{\rm part} \right>$ values.  The values of ratios, 
integrated in the high-$p_T$ region 
[Fig.~\ref{fig:IntegratedRatios}(b)~and~\ref{fig:IntegratedRatios}(d)], 
smoothly grow with increasing $\left< N_{\rm part} \right>$. Because the 
$\bar{p}/p$ ratio is approximately equal to 0.73~\cite{PPG026}, 
regardless of $\left<N_{\rm part}\right>$ the integrated 
$p/\pi^{+}$ ratios exceed the integrated $\bar{p}/\pi^{-}$ 
ratios.

According to the recombination model, the $p_T$ of a produced 
hadron is the sum of the $p_T$ of its constituent quarks. 
Therefore, baryon-invariant $p_T$ spectra are shifted towards 
larger $p_T$ relative to the meson $p_T$ spectra, which leads to 
an enhancement in baryon over meson production in the 
intermediate $p_T$ region. Therefore, the behavior of the 
integrated ratio supports the assumption that the influence of 
the recombination mechanism grows with increasing 
$\left<N_{\rm part}\right>$.



\section{SUMMARY}

The PHENIX experiment has measured identified charged-hadron invariant 
\pt and \mt spectra and nuclear-modification factors \rab, and $p/\pi$, 
$K/\pi$ ratios in \pal, \heau, \cuau collisions at \sqsn = 200 GeV and 
in \uu collisions at \sqsn = 193 GeV.

The values of freeze-out temperatures \To and average collective 
velocities \ut have been obtained. The \To values do not exhibit any 
dependence on the collision centrality and \Npart values, whereas the 
values of \ut smoothly increase with increasing of \Npart values. This 
indicates that in collisions characterized by large \Npart values 
(central \cuau, \auau, \uu collisions), collective effects are more 
pronounced than in collision systems with small \Npart values (\pal, 
\heau collisions and peripheral collisions of large systems).

The \prot/$\pi$ and $K/\pi$ ratios have been measured to investigate 
differences in baryon vs.~meson production in large and small collision 
systems. In collisions with small \Npart values the ratios of 
\prot/$\pi$ are comparable to those measured in \pp collisions 
($(p/\pi)^{p+p}$). In collision systems with large \Npart values 
\prot/$\pi$ ratios reach the values of $\approx\,0.6$, which is $\approx\,2$ 
times larger than $(p/\pi)^{p+p}$.

In heavy ion collisions (\cuau and \uu) the \prot/$\pi$ ratios exhibit 
strong centrality dependence, but in small \heau collisions the 
centrality dependence is much more modest due to the much smaller range 
of \Npart values.  Within uncertainties, the \prot/$\pi$ values measured 
in each centrality of \pal collisions are consistent with those measured 
in \pp collisions. The values of the $K/\pi$ ratios show a modest 
centrality dependence, which is insignificant within systematic 
uncertainties. The observed behavior of $p/\pi$ and $K/\pi$ ratios can 
be qualitatively described through hadronization by 
recombination~\cite{Recombination1,Recombination2}.

Comparison of identified charged-hadron nuclear-modification factors 
shows that in \dau, \heau, Cu$+$Au, Au$+$Au, and U$+$U collision systems the 
\rab values are consistent at the same number of participant nucleons 
\Npart. This indicates that identified charged-hadron production does 
not depend on the geometry and collision species, but rather is 
determined by system size alone (as indicated by \Npart values). Further, 
it is found that: (i) the slope of $R_{AB}(p_T)$ in \pal collisions is 
flatter than in \heau and \dau collisions at the same \Npart values, 
(ii) proton \rab values in \pal collisions are equal to unity in the 
range of 1.0 GeV/$c$ $< p_T <$ 2.5 GeV/$c$, while proton \rab values 
measured in \heau and \dau collisions are larger than unity. Despite the 
observed absence of proton enhancement in \pal collisions, there were 
evidences of QGP formation found by the PHENIX experiment in studies of 
$J/\psi$, $\psi(2S)$, and charged-hadron production at backward 
rapidity~\cite{psi_SmallSyst}. The observed differences in identified charged 
hadron production in \pal and $d$/\heau collisions may be caused by the 
size of the system created in \pal collisions being too small to observe 
the expected increase in proton production.

\section*{ACKNOWLEDGMENTS}

We thank the staff of the Collider-Accelerator and Physics
Departments at Brookhaven National Laboratory and the staff of
the other PHENIX participating institutions for their vital
contributions.
We acknowledge support from the Office of Nuclear Physics in the
Office of Science of the Department of Energy,
the National Science Foundation,
Abilene Christian University Research Council,
Research Foundation of SUNY, and
Dean of the College of Arts and Sciences, Vanderbilt University
(U.S.A),
Ministry of Education, Culture, Sports, Science, and Technology
and the Japan Society for the Promotion of Science (Japan),
Natural Science Foundation of China (People's Republic of China),
Croatian Science Foundation and
Ministry of Science and Education (Croatia),
Ministry of Education, Youth and Sports (Czech Republic),
Centre National de la Recherche Scientifique, Commissariat
{\`a} l'{\'E}nergie Atomique, and Institut National de Physique
Nucl{\'e}aire et de Physique des Particules (France),
J. Bolyai Research Scholarship, EFOP, HUN-REN ATOMKI, NKFIH,
and OTKA (Hungary),
Department of Atomic Energy and Department of Science and Technology
(India),
Israel Science Foundation (Israel),
Basic Science Research and SRC(CENuM) Programs through NRF
funded by the Ministry of Education and the Ministry of
Science and ICT (Korea).
Ministry of Education and Science, Russian Academy of Sciences,
Federal Agency of Atomic Energy (Russia),
VR and Wallenberg Foundation (Sweden),
University of Zambia, the Government of the Republic of Zambia (Zambia),
the U.S. Civilian Research and Development Foundation for the
Independent States of the Former Soviet Union,
the Hungarian American Enterprise Scholarship Fund,
the US-Hungarian Fulbright Foundation,
and the US-Israel Binational Science Foundation.


\begin{thebibliography}{48}%
\makeatletter
\providecommand \@ifxundefined [1]{%
 \@ifx{#1\undefined}
}%
\providecommand \@ifnum [1]{%
 \ifnum #1\expandafter \@firstoftwo
 \else \expandafter \@secondoftwo
 \fi
}%
\providecommand \@ifx [1]{%
 \ifx #1\expandafter \@firstoftwo
 \else \expandafter \@secondoftwo
 \fi
}%
\providecommand \natexlab [1]{#1}%
\providecommand \enquote  [1]{``#1''}%
\providecommand \bibnamefont  [1]{#1}%
\providecommand \bibfnamefont [1]{#1}%
\providecommand \citenamefont [1]{#1}%
\providecommand \href@noop [0]{\@secondoftwo}%
\providecommand \href [0]{\begingroup \@sanitize@url \@href}%
\providecommand \@href[1]{\@@startlink{#1}\@@href}%
\providecommand \@@href[1]{\endgroup#1\@@endlink}%
\providecommand \@sanitize@url [0]{\catcode `\\12\catcode `\$12\catcode
  `\&12\catcode `\#12\catcode `\^12\catcode `\_12\catcode `\%12\relax}%
\providecommand \@@startlink[1]{}%
\providecommand \@@endlink[0]{}%
\providecommand \url  [0]{\begingroup\@sanitize@url \@url }%
\providecommand \@url [1]{\endgroup\@href {#1}{\urlprefix }}%
\providecommand \urlprefix  [0]{URL }%
\providecommand \Eprint [0]{\href }%
\providecommand \doibase [0]{https://doi.org/}%
\providecommand \selectlanguage [0]{\@gobble}%
\providecommand \bibinfo  [0]{\@secondoftwo}%
\providecommand \bibfield  [0]{\@secondoftwo}%
\providecommand \translation [1]{[#1]}%
\providecommand \BibitemOpen [0]{}%
\providecommand \bibitemStop [0]{}%
\providecommand \bibitemNoStop [0]{.\EOS\space}%
\providecommand \EOS [0]{\spacefactor3000\relax}%
\providecommand \BibitemShut  [1]{\csname bibitem#1\endcsname}%
\let\auto@bib@innerbib\@empty
\bibitem [{\citenamefont {Adcox}\ \emph {et~al.}(2005)\citenamefont {Adcox}
  \emph {et~al.}}]{QGP}%
  \BibitemOpen
  \bibfield  {author} {\bibinfo {author} {\bibfnamefont {K.}~\bibnamefont
  {Adcox}} \emph {et~al.} (\bibinfo {collaboration} {PHENIX Collaboration}),\
  }\bibfield  {title} {\bibinfo {title} {{Formation of dense partonic matter in
  relativistic nucleus-nucleus collisions at RHIC: Experimental evaluation by
  the PHENIX Collaboration}},\ }\href@noop {} {\bibfield  {journal} {\bibinfo
  {journal} {Nucl. Phys. A}\ }\textbf {\bibinfo {volume} {757}},\ \bibinfo
  {pages} {184} (\bibinfo {year} {2005})}\BibitemShut {NoStop}%
\bibitem [{\citenamefont {Adams}\ \emph {et~al.}(2005)\citenamefont {Adams}
  \emph {et~al.}}]{QGP_STAR}%
  \BibitemOpen
  \bibfield  {author} {\bibinfo {author} {\bibfnamefont {J.}~\bibnamefont
  {Adams}} \emph {et~al.} (\bibinfo {collaboration} {{STAR Collaboration}}),\
  }\bibfield  {title} {\bibinfo {title} {{Experimental and theoretical
  challenges in the search for the quark gluon plasma: The STAR Collaboration's
  critical assessment of the evidence from RHIC collisions}},\ }\href@noop {}
  {\bibfield  {journal} {\bibinfo  {journal} {Nucl. Phys. A}\ }\textbf
  {\bibinfo {volume} {757}},\ \bibinfo {pages} {102} (\bibinfo {year}
  {2005})}\BibitemShut {NoStop}%
\bibitem [{\citenamefont {Niida}\ and\ \citenamefont
  {Miake}()}]{QGP_signatures}%
  \BibitemOpen
  \bibfield  {author} {\bibinfo {author} {\bibfnamefont {T.}~\bibnamefont
  {Niida}}\ and\ \bibinfo {author} {\bibfnamefont {Y.}~\bibnamefont {Miake}},\
  }\href@noop {} {\bibinfo {title} {{Signatures of QGP at RHIC and the LHC}}},\
  \bibinfo {note} {{arXiv:2104.11406}}\BibitemShut {NoStop}%
\bibitem [{\citenamefont {Koch}\ \emph {et~al.}(1986)\citenamefont {Koch},
  \citenamefont {Muller},\ and\ \citenamefont
  {Rafelski}}]{StrangenessEnhancenment}%
  \BibitemOpen
  \bibfield  {author} {\bibinfo {author} {\bibfnamefont {P.}~\bibnamefont
  {Koch}}, \bibinfo {author} {\bibfnamefont {B.}~\bibnamefont {Muller}},\ and\
  \bibinfo {author} {\bibfnamefont {J.}~\bibnamefont {Rafelski}},\ }\bibfield
  {title} {\bibinfo {title} {{Strangeness in Relativistic Heavy Ion
  Collisions}},\ }\href@noop {} {\bibfield  {journal} {\bibinfo  {journal}
  {Phys. Rept.}\ }\textbf {\bibinfo {volume} {142}},\ \bibinfo {pages} {167}
  (\bibinfo {year} {1986})}\BibitemShut {NoStop}%
\bibitem [{\citenamefont {Koch}\ \emph {et~al.}(2017)\citenamefont {Koch},
  \citenamefont {M\"uller},\ and\ \citenamefont {Rafelski}}]{Strangeness_QGP}%
  \BibitemOpen
  \bibfield  {author} {\bibinfo {author} {\bibfnamefont {P.}~\bibnamefont
  {Koch}}, \bibinfo {author} {\bibfnamefont {B.}~\bibnamefont {M\"uller}},\
  and\ \bibinfo {author} {\bibfnamefont {J.}~\bibnamefont {Rafelski}},\
  }\bibfield  {title} {\bibinfo {title} {{From Strangeness Enhancement to
  Quark-Gluon Plasma Discovery}},\ }\href@noop {} {\bibfield  {journal}
  {\bibinfo  {journal} {Int. J. Mod. Phys. A}\ }\textbf {\bibinfo {volume}
  {32}},\ \bibinfo {pages} {1730024} (\bibinfo {year} {2017})}\BibitemShut
  {NoStop}%
\bibitem [{\citenamefont {J.}\ \emph {et~al.}(2023)\citenamefont {J.} \emph
  {et~al.}}]{phi_LargeSysts}%
  \BibitemOpen
  \bibfield  {author} {\bibinfo {author} {\bibfnamefont {A.~N.}\ \bibnamefont
  {J.}} \emph {et~al.} (\bibinfo {collaboration} {PHENIX Collaboration}),\
  }\bibfield  {title} {\bibinfo {title} {{Measurement of $\phi$-meson
  production in Cu$+$Au collisions at $\sqrt{s_{_{NN}}}=200$ GeV and U$+$U
  collisions at $\sqrt{s_{_{NN}}}=193$ GeV}},\ }\href@noop {} {\bibfield
  {journal} {\bibinfo  {journal} {Phys. Rev. C}\ }\textbf {\bibinfo {volume}
  {107}},\ \bibinfo {pages} {014907} (\bibinfo {year} {2023})}\BibitemShut
  {NoStop}%
\bibitem [{\citenamefont {Bellini}(2017)}]{ALICE_strangeness}%
  \BibitemOpen
  \bibfield  {author} {\bibinfo {author} {\bibfnamefont {F.}~\bibnamefont
  {Bellini}} (\bibinfo {collaboration} {{ALICE Collaboration}}),\ }\bibfield
  {title} {\bibinfo {title} {Strangeness in {ALICE at the LHC}},\ }\href@noop
  {} {\bibfield  {journal} {\bibinfo  {journal} {J. Phys.: Conf. Series}\
  }\textbf {\bibinfo {volume} {779}},\ \bibinfo {pages} {012007} (\bibinfo
  {year} {2017})}\BibitemShut {NoStop}%
\bibitem [{\citenamefont {Shi}(2017)}]{STAR_strangeness}%
  \BibitemOpen
  \bibfield  {author} {\bibinfo {author} {\bibfnamefont {S.}~\bibnamefont
  {Shi}} (\bibinfo {collaboration} {{STAR collaboration}}),\ }\bibfield
  {title} {\bibinfo {title} {{Strangeness in STAR experiment at RHIC}},\ }\href
  {https://doi.org/10.1088/1742-6596/779/1/012008} {\bibfield  {journal}
  {\bibinfo  {journal} {J. Phys.: Conf. Series}\ }\textbf {\bibinfo {volume}
  {779}},\ \bibinfo {pages} {012008} (\bibinfo {year} {2017})}\BibitemShut
  {NoStop}%
\bibitem [{\citenamefont {Rohrich}\ \emph {et~al.}(2005)\citenamefont {Rohrich}
  \emph {et~al.}}]{BRAHMS_Strangeness_CH}%
  \BibitemOpen
  \bibfield  {author} {\bibinfo {author} {\bibfnamefont {D.}~\bibnamefont
  {Rohrich}} \emph {et~al.} (\bibinfo {collaboration} {{BRAHMS
  collaboration}}),\ }\bibfield  {title} {\bibinfo {title} {{Strangeness
  production at RHIC: recent results from BRAHMS}},\ }\href@noop {} {\bibfield
  {journal} {\bibinfo  {journal} {J. Phys. G: Nucl. Part. Phys.}\ }\textbf
  {\bibinfo {volume} {31}},\ \bibinfo {pages} {S65} (\bibinfo {year}
  {2005})}\BibitemShut {NoStop}%
\bibitem [{\citenamefont {d'Enterria}(2010)}]{JetQuenching1}%
  \BibitemOpen
  \bibfield  {author} {\bibinfo {author} {\bibfnamefont {D.}~\bibnamefont
  {d'Enterria}},\ }\bibfield  {title} {\bibinfo {title} {{Jet quenching}},\
  }\href@noop {} {\bibfield  {journal} {\bibinfo  {journal} {Landolt-Bernstein
 - Group I}\ ,\ \bibinfo {pages} {471}} (\bibinfo {year} {2010})}\BibitemShut
  {NoStop}%
\bibitem [{\citenamefont {Cao}\ and\ \citenamefont
  {Wang}(2021)}]{JetQuenching2}%
  \BibitemOpen
  \bibfield  {author} {\bibinfo {author} {\bibfnamefont {S.}~\bibnamefont
  {Cao}}\ and\ \bibinfo {author} {\bibfnamefont {X.-N.}\ \bibnamefont {Wang}},\
  }\bibfield  {title} {\bibinfo {title} {{Jet quenching and medium response in
  high-energy heavy-ion collisions: A review}},\ }\href@noop {} {\bibfield
  {journal} {\bibinfo  {journal} {Reports Prog. Phys.}\ }\textbf {\bibinfo
  {volume} {84}},\ \bibinfo {pages} {024301} (\bibinfo {year}
  {2021})}\BibitemShut {NoStop}%
\bibitem [{\citenamefont {L{\'{e}}vai}\ \emph {et~al.}(2002)\citenamefont
  {L{\'{e}}vai}, \citenamefont {Papp}, \citenamefont {Fai}, \citenamefont
  {Gyulassy}, \citenamefont {Barnafoldi}, \citenamefont {Vitev},\ and\
  \citenamefont {Zhang}}]{JetQuenching3}%
  \BibitemOpen
  \bibfield  {author} {\bibinfo {author} {\bibfnamefont {P.}~\bibnamefont
  {L{\'{e}}vai}}, \bibinfo {author} {\bibfnamefont {G.}~\bibnamefont {Papp}},
  \bibinfo {author} {\bibfnamefont {G.}~\bibnamefont {Fai}}, \bibinfo {author}
  {\bibfnamefont {M.}~\bibnamefont {Gyulassy}}, \bibinfo {author}
  {\bibfnamefont {G.}~\bibnamefont {Barnafoldi}}, \bibinfo {author}
  {\bibfnamefont {I.}~\bibnamefont {Vitev}},\ and\ \bibinfo {author}
  {\bibfnamefont {Y.}~\bibnamefont {Zhang}},\ }\bibfield  {title} {\bibinfo
  {title} {{Discovery of jet quenching at RHIC and the opacity of the produced
  gluon plasma}},\ }\href@noop {} {\bibfield  {journal} {\bibinfo  {journal}
  {Nucl. Phys. A}\ }\textbf {\bibinfo {volume} {698}},\ \bibinfo {pages} {631}
  (\bibinfo {year} {2002})}\BibitemShut {NoStop}%
\bibitem [{\citenamefont {Reed}(2015)}]{ALICE_jet_quenching}%
  \BibitemOpen
  \bibfield  {author} {\bibinfo {author} {\bibfnamefont {R.}~\bibnamefont
  {Reed}} (\bibinfo {collaboration} {{ALICE Collaboration}}),\ }\bibfield
  {title} {\bibinfo {title} {{Jet production in pp, pPb and PbPb collisions
  measured by ALICE}},\ }\href@noop {} {\bibfield  {journal} {\bibinfo
  {journal} {J. Phys.: Conf. Series}\ }\textbf {\bibinfo {volume} {636}},\
  \bibinfo {pages} {012010} (\bibinfo {year} {2015})}\BibitemShut {NoStop}%
\bibitem [{\citenamefont {Adler}\ \emph
  {et~al.}(2004{\natexlab{a}})\citenamefont {Adler} \emph {et~al.}}]{PPG026}%
  \BibitemOpen
  \bibfield  {author} {\bibinfo {author} {\bibfnamefont {S.~S.}\ \bibnamefont
  {Adler}} \emph {et~al.} (\bibinfo {collaboration} {PHENIX Collaboration}),\
  }\bibfield  {title} {\bibinfo {title} {{Identified charged particle spectra
  and yields in Au+Au collisions at $\sqrt{s_{NN}}$ = 200 GeV}},\ }\href@noop
  {} {\bibfield  {journal} {\bibinfo  {journal} {Phys. Rev. C}\ }\textbf
  {\bibinfo {volume} {69}},\ \bibinfo {pages} {034909} (\bibinfo {year}
  {2004}{\natexlab{a}})}\BibitemShut {NoStop}%
\bibitem [{\citenamefont {Adcox}\ \emph {et~al.}(2002)\citenamefont {Adcox}
  \emph {et~al.}}]{CHspectra_130GeV}%
  \BibitemOpen
  \bibfield  {author} {\bibinfo {author} {\bibfnamefont {K.}~\bibnamefont
  {Adcox}} \emph {et~al.} (\bibinfo {collaboration} {PHENIX}),\ }\bibfield
  {title} {\bibinfo {title} {{Centrality dependence of $\pi^+ / \pi^-, K^+ /
  K^-, p$ and anti-$p$ production from $\sqrt{s}_{NN}=130$ GeV Au+Au collisions
  at RHIC}},\ }\href@noop {} {\bibfield  {journal} {\bibinfo  {journal} {Phys.
  Rev. Lett.}\ }\textbf {\bibinfo {volume} {88}},\ \bibinfo {pages} {242301}
  (\bibinfo {year} {2002})}\BibitemShut {NoStop}%
\bibitem [{\citenamefont {Velkovska}(2002)}]{Velkovska_2002}%
  \BibitemOpen
  \bibfield  {author} {\bibinfo {author} {\bibfnamefont {J.}~\bibnamefont
  {Velkovska}},\ }\bibfield  {title} {\bibinfo {title} {{$p_T$ distributions of
  identified charged hadrons measured with the PHENIX experiment at RHIC}},\
  }\href@noop {} {\bibfield  {journal} {\bibinfo  {journal} {Nucl. Phys. A}\
  }\textbf {\bibinfo {volume} {698}},\ \bibinfo {pages} {507} (\bibinfo {year}
  {2002})}\BibitemShut {NoStop}%
\bibitem [{\citenamefont {Adare}\ \emph {et~al.}(2013)\citenamefont {Adare}
  \emph {et~al.}}]{ppg146}%
  \BibitemOpen
  \bibfield  {author} {\bibinfo {author} {\bibfnamefont {A.}~\bibnamefont
  {Adare}} \emph {et~al.} (\bibinfo {collaboration} {PHENIX Collaboration}),\
  }\bibfield  {title} {\bibinfo {title} {{Spectra and ratios of identified
  particles in Au+Au and $d$+Au collisions at $\sqrt{s_{NN}}=200$ GeV}},\
  }\href@noop {} {\bibfield  {journal} {\bibinfo  {journal} {Phys. Rev. C}\
  }\textbf {\bibinfo {volume} {88}},\ \bibinfo {pages} {024906} (\bibinfo
  {year} {2013})}\BibitemShut {NoStop}%
\bibitem [{\citenamefont {Back}\ \emph {et~al.}(2007)\citenamefont {Back} \emph
  {et~al.}}]{PHOBOS_CH}%
  \BibitemOpen
  \bibfield  {author} {\bibinfo {author} {\bibfnamefont {B.~B.}\ \bibnamefont
  {Back}} \emph {et~al.} (\bibinfo {collaboration} {{PHOBOS Collaboration}}),\
  }\bibfield  {title} {\bibinfo {title} {{Identified hadron transverse momentum
  spectra in Au+Au collisions at $\sqrt{s_{NN}}$ = 62.4 GeV}},\ }\href@noop {}
  {\bibfield  {journal} {\bibinfo  {journal} {Phys. Rev. C}\ }\textbf {\bibinfo
  {volume} {75}},\ \bibinfo {pages} {024910} (\bibinfo {year}
  {2007})}\BibitemShut {NoStop}%
\bibitem [{\citenamefont {Adams}\ \emph {et~al.}(2006)\citenamefont {Adams}
  \emph {et~al.}}]{STAR_CH}%
  \BibitemOpen
  \bibfield  {author} {\bibinfo {author} {\bibfnamefont {J.}~\bibnamefont
  {Adams}} \emph {et~al.} (\bibinfo {collaboration} {{STAR collaboration}}),\
  }\bibfield  {title} {\bibinfo {title} {{Identified hadron spectra at large
  transverse momentum in p+p and d+Au collisions at $\sqrt{s_{NN}}$=200 GeV}},\
  }\href@noop {} {\bibfield  {journal} {\bibinfo  {journal} {Phys. Lett. B}\
  }\textbf {\bibinfo {volume} {637}},\ \bibinfo {pages} {161} (\bibinfo {year}
  {2006})}\BibitemShut {NoStop}%
\bibitem [{\citenamefont {Fries}\ \emph {et~al.}(2008)\citenamefont {Fries},
  \citenamefont {Greco},\ and\ \citenamefont {Sorensen}}]{Coalescence_models}%
  \BibitemOpen
  \bibfield  {author} {\bibinfo {author} {\bibfnamefont {R.}~\bibnamefont
  {Fries}}, \bibinfo {author} {\bibfnamefont {V.}~\bibnamefont {Greco}},\ and\
  \bibinfo {author} {\bibfnamefont {P.}~\bibnamefont {Sorensen}},\ }\bibfield
  {title} {\bibinfo {title} {{Coalescence Models for Hadron Formation from
  Quark-Gluon Plasma}},\ }\href
  {https://doi.org/10.1146/annurev.nucl.58.110707.171134} {\bibfield  {journal}
  {\bibinfo  {journal} {Ann. Rev. Nucl. Part. Sci}\ }\textbf {\bibinfo {volume}
  {58}},\ \bibinfo {pages} {177} (\bibinfo {year} {2008})}\BibitemShut
  {NoStop}%
\bibitem [{\citenamefont {Kumar}(2011)}]{STAR_UU}%
  \BibitemOpen
  \bibfield  {author} {\bibinfo {author} {\bibfnamefont {L.}~\bibnamefont
  {Kumar}} (\bibinfo {collaboration} {{for the STAR collaboration}}),\
  }\bibfield  {title} {\bibinfo {title} {{Identified hadron production from the
  RHIC beam energy scan}},\ }\href@noop {} {\bibfield  {journal} {\bibinfo
  {journal} {J. Phys. G Nucl. Part. Phys.}\ }\textbf {\bibinfo {volume} {38}},\
  \bibinfo {pages} {124145} (\bibinfo {year} {2011})}\BibitemShut {NoStop}%
\bibitem [{\citenamefont {Adcox}\ \emph
  {et~al.}(2003{\natexlab{a}})\citenamefont {Adcox} \emph
  {et~al.}}]{PHENIXoverview}%
  \BibitemOpen
  \bibfield  {author} {\bibinfo {author} {\bibfnamefont {K.}~\bibnamefont
  {Adcox}} \emph {et~al.} (\bibinfo {collaboration} {PHENIX Collaboration}),\
  }\bibfield  {title} {\bibinfo {title} {{PHENIX detector overview}},\
  }\href@noop {} {\bibfield  {journal} {\bibinfo  {journal} {Nucl. Instrum.
  Methods Phys. Res., Sec. A}\ }\textbf {\bibinfo {volume} {499}},\ \bibinfo
  {pages} {469} (\bibinfo {year} {2003}{\natexlab{a}})}\BibitemShut {NoStop}%
\bibitem [{\citenamefont {Greco}\ \emph
  {et~al.}(2003{\natexlab{a}})\citenamefont {Greco}, \citenamefont {Ko},\ and\
  \citenamefont {L\'evai}}]{Recombination1}%
  \BibitemOpen
  \bibfield  {author} {\bibinfo {author} {\bibfnamefont {V.}~\bibnamefont
  {Greco}}, \bibinfo {author} {\bibfnamefont {C.~M.}\ \bibnamefont {Ko}},\ and\
  \bibinfo {author} {\bibfnamefont {P.}~\bibnamefont {L\'evai}},\ }\bibfield
  {title} {\bibinfo {title} {{Parton Coalescence and the Antiproton/Pion
  Anomaly at RHIC}},\ }\href@noop {} {\bibfield  {journal} {\bibinfo  {journal}
  {Phys. Rev. Lett.}\ }\textbf {\bibinfo {volume} {90}},\ \bibinfo {pages}
  {202302} (\bibinfo {year} {2003}{\natexlab{a}})}\BibitemShut {NoStop}%
\bibitem [{\citenamefont {Hwa}\ and\ \citenamefont
  {Yang}(2003)}]{Recombination2}%
  \BibitemOpen
  \bibfield  {author} {\bibinfo {author} {\bibfnamefont {R.~C.}\ \bibnamefont
  {Hwa}}\ and\ \bibinfo {author} {\bibfnamefont {C.~B.}\ \bibnamefont {Yang}},\
  }\bibfield  {title} {\bibinfo {title} {{Scaling behavior at high ${p}_{T}$
  and the $p/\ensuremath{\pi}$ ratio}},\ }\href@noop {} {\bibfield  {journal}
  {\bibinfo  {journal} {Phys. Rev. C}\ }\textbf {\bibinfo {volume} {67}},\
  \bibinfo {pages} {034902} (\bibinfo {year} {2003})}\BibitemShut {NoStop}%
\bibitem [{\citenamefont {Dorso}\ \emph {et~al.}()\citenamefont {Dorso},
  \citenamefont {Molinelli}, \citenamefont {Nichols},\ and\ \citenamefont
  {Lopez}}]{CNM}%
  \BibitemOpen
  \bibfield  {author} {\bibinfo {author} {\bibfnamefont {C.~O.}\ \bibnamefont
  {Dorso}}, \bibinfo {author} {\bibfnamefont {P.~A.~G.}\ \bibnamefont
  {Molinelli}}, \bibinfo {author} {\bibfnamefont {J.~I.}\ \bibnamefont
  {Nichols}},\ and\ \bibinfo {author} {\bibfnamefont {J.~A.}\ \bibnamefont
  {Lopez}},\ }\href@noop {} {\bibinfo {title} {{Cold nuclear matter}}},\
  \bibinfo {note} {arXiv:1211.5582}\BibitemShut {NoStop}%
\bibitem [{\citenamefont {Pal}(2005)}]{phi_dAu}%
  \BibitemOpen
  \bibfield  {author} {\bibinfo {author} {\bibfnamefont {D.}~\bibnamefont
  {Pal}} (\bibinfo {collaboration} {{for the PHENIX Collaboration}}),\
  }\bibfield  {title} {\bibinfo {title} {{$\phi$ meson production in d$+$Au
  collisions at $\sqrt{s_{NN}}=200$ GeV}},\ }\href@noop {} {\bibfield
  {journal} {\bibinfo  {journal} {J. Phys. G}\ }\textbf {\bibinfo {volume}
  {31}},\ \bibinfo {pages} {S211} (\bibinfo {year} {2005})}\BibitemShut
  {NoStop}%
\bibitem [{\citenamefont {Krelina}\ and\ \citenamefont
  {Nemchik}(2014)}]{Cronin}%
  \BibitemOpen
  \bibfield  {author} {\bibinfo {author} {\bibfnamefont {M.}~\bibnamefont
  {Krelina}}\ and\ \bibinfo {author} {\bibfnamefont {J.}~\bibnamefont
  {Nemchik}},\ }\bibfield  {title} {\bibinfo {title} {{Cronin effect at
  different energies: from RHIC to LHC}},\ }\href@noop {} {\bibfield  {journal}
  {\bibinfo  {journal} {EPJ Web of Conf.}\ }\textbf {\bibinfo {volume} {66}}
  (\bibinfo {year} {2014})}\BibitemShut {NoStop}%
\bibitem [{\citenamefont {Shao}(2006)}]{Cronin_STAR}%
  \BibitemOpen
  \bibfield  {author} {\bibinfo {author} {\bibfnamefont {M.}~\bibnamefont
  {Shao}} (\bibinfo {collaboration} {{STAR collaboration}}),\ }\bibfield
  {title} {\bibinfo {title} {{Cronin effect at RHIC}},\ }\href@noop {}
  {\bibfield  {journal} {\bibinfo  {journal} {AIP Conf. Proc.}\ }\textbf
  {\bibinfo {volume} {828}},\ \bibinfo {pages} {49} (\bibinfo {year}
  {2006})}\BibitemShut {NoStop}%
\bibitem [{\citenamefont {Wang}\ and\ \citenamefont {Wang}(2002)}]{MPI2}%
  \BibitemOpen
  \bibfield  {author} {\bibinfo {author} {\bibfnamefont {E.}~\bibnamefont
  {Wang}}\ and\ \bibinfo {author} {\bibfnamefont {X.-N.}\ \bibnamefont
  {Wang}},\ }\bibfield  {title} {\bibinfo {title} {{Jet Tomography of Hot and
  Cold Nuclear Matter}},\ }\href@noop {} {\bibfield  {journal} {\bibinfo
  {journal} {Phys. Rev. Lett.}\ }\textbf {\bibinfo {volume} {89}},\ \bibinfo
  {pages} {162301} (\bibinfo {year} {2002})}\BibitemShut {NoStop}%
\bibitem [{\citenamefont {Kova{\v{r}}{\'{\i}}k}\ \emph
  {et~al.}(2016)\citenamefont {Kova{\v{r}}{\'{\i}}k}, \citenamefont {Kusina},
  \citenamefont {Je{\v{z}}o}, \citenamefont {Clark}, \citenamefont {Keppel},
  \citenamefont {Lyonnet}, \citenamefont {Morf{\'{\i}}n}, \citenamefont
  {Olness}, \citenamefont {Owens}, \citenamefont {Schienbein},\ and\
  \citenamefont {Yu}}]{PDF1}%
  \BibitemOpen
  \bibfield  {author} {\bibinfo {author} {\bibfnamefont {K.}~\bibnamefont
  {Kova{\v{r}}{\'{\i}}k}}, \bibinfo {author} {\bibfnamefont {A.}~\bibnamefont
  {Kusina}}, \bibinfo {author} {\bibfnamefont {T.}~\bibnamefont {Je{\v{z}}o}},
  \bibinfo {author} {\bibfnamefont {D.~B.}\ \bibnamefont {Clark}}, \bibinfo
  {author} {\bibfnamefont {C.}~\bibnamefont {Keppel}}, \bibinfo {author}
  {\bibfnamefont {F.}~\bibnamefont {Lyonnet}}, \bibinfo {author} {\bibfnamefont
  {J.~G.}\ \bibnamefont {Morf{\'{\i}}n}}, \bibinfo {author} {\bibfnamefont
  {F.~I.}\ \bibnamefont {Olness}}, \bibinfo {author} {\bibfnamefont {J.~F.}\
  \bibnamefont {Owens}}, \bibinfo {author} {\bibfnamefont {I.}~\bibnamefont
  {Schienbein}},\ and\ \bibinfo {author} {\bibfnamefont {J.~Y.}\ \bibnamefont
  {Yu}},\ }\bibfield  {title} {\bibinfo {title} {{nCTEQ15: Global analysis of
  nuclear parton distributions with uncertainties in the CTEQ framework}},\
  }\href@noop {} {\bibfield  {journal} {\bibinfo  {journal} {Phys. Rev. D}\
  }\textbf {\bibinfo {volume} {93}},\ \bibinfo {pages} {085037} (\bibinfo
  {year} {2016})}\BibitemShut {NoStop}%
\bibitem [{\citenamefont {Abdallah}\ \emph {et~al.}(2022)\citenamefont
  {Abdallah} \emph {et~al.}}]{PDF2}%
  \BibitemOpen
  \bibfield  {author} {\bibinfo {author} {\bibfnamefont {M.}~\bibnamefont
  {Abdallah}} \emph {et~al.},\ }\bibfield  {title} {\bibinfo {title}
  {{Measurement of cold nuclear matter effects for inclusive {$J/\psi$} in
  {$p+Au$} collisions at $\sqrt{s_{NN}}$=200 {GeV}}},\ }\href@noop {}
  {\bibfield  {journal} {\bibinfo  {journal} {Phys. Lett. B}\ }\textbf
  {\bibinfo {volume} {825}},\ \bibinfo {pages} {136865} (\bibinfo {year}
  {2022})}\BibitemShut {NoStop}%
\bibitem [{\citenamefont {Aidala}\ \emph {et~al.}(2019)\citenamefont {Aidala}
  \emph {et~al.}}]{PHENIX_Nature}%
  \BibitemOpen
  \bibfield  {author} {\bibinfo {author} {\bibfnamefont {C.}~\bibnamefont
  {Aidala}} \emph {et~al.} (\bibinfo {collaboration} {PHENIX Collaboration}),\
  }\bibfield  {title} {\bibinfo {title} {{Creation of quark-gluon plasma
  droplets with three distinct geometries}},\ }\href@noop {} {\bibfield
  {journal} {\bibinfo  {journal} {Nature Phys.}\ }\textbf {\bibinfo {volume}
  {15}},\ \bibinfo {pages} {214} (\bibinfo {year} {2019})}\BibitemShut
  {NoStop}%
\bibitem [{\citenamefont {Adcox}\ \emph
  {et~al.}(2003{\natexlab{b}})\citenamefont {Adcox} \emph
  {et~al.}}]{TrackingSystem}%
  \BibitemOpen
  \bibfield  {author} {\bibinfo {author} {\bibfnamefont {K.}~\bibnamefont
  {Adcox}} \emph {et~al.} (\bibinfo {collaboration} {PHENIX Collaboration}),\
  }\bibfield  {title} {\bibinfo {title} {{PHENIX central arm tracking
  detectors}},\ }\href@noop {} {\bibfield  {journal} {\bibinfo  {journal}
  {Nucl. Instrum. Methods Phys. Res., Sec. A}\ }\textbf {\bibinfo {volume}
  {499}},\ \bibinfo {pages} {489} (\bibinfo {year}
  {2003}{\natexlab{b}})}\BibitemShut {NoStop}%
\bibitem [{\citenamefont {Carl\'en}\ \emph {et~al.}(1999)\citenamefont
  {Carl\'en} \emph {et~al.}}]{ToF}%
  \BibitemOpen
  \bibfield  {author} {\bibinfo {author} {\bibfnamefont {L.}~\bibnamefont
  {Carl\'en}} \emph {et~al.},\ }\bibfield  {title} {\bibinfo {title} {{A
  large-acceptance spectrometer for tracking in a high multiplicity
  environment, based on space point measurements and high resolution
  time-of-flight}},\ }\href@noop {} {\bibfield  {journal} {\bibinfo  {journal}
  {Nucl. Instrum. Methods Phys. Res., Sec. A}\ }\textbf {\bibinfo {volume}
  {431}},\ \bibinfo {pages} {123} (\bibinfo {year} {1999})}\BibitemShut
  {NoStop}%
\bibitem [{\citenamefont {Allen}\ \emph {et~al.}(2003)\citenamefont {Allen}
  \emph {et~al.}}]{PHENIX_InnerDetectors}%
  \BibitemOpen
  \bibfield  {author} {\bibinfo {author} {\bibfnamefont {M.}~\bibnamefont
  {Allen}} \emph {et~al.} (\bibinfo {collaboration} {PHENIX Collaboration}),\
  }\bibfield  {title} {\bibinfo {title} {{PHENIX inner detectors}},\
  }\href@noop {} {\bibfield  {journal} {\bibinfo  {journal} {Nucl. Instrum.
  Methods Phys. Res., Sec. A}\ }\textbf {\bibinfo {volume} {499}},\ \bibinfo
  {pages} {549} (\bibinfo {year} {2003})}\BibitemShut {NoStop}%
\bibitem [{\citenamefont {Aizawa}\ \emph {et~al.}(2003)\citenamefont {Aizawa}
  \emph {et~al.}}]{PHENIX_CAdetectors}%
  \BibitemOpen
  \bibfield  {author} {\bibinfo {author} {\bibfnamefont {M.}~\bibnamefont
  {Aizawa}} \emph {et~al.} (\bibinfo {collaboration} {PHENIX Collaboration}),\
  }\bibfield  {title} {\bibinfo {title} {{PHENIX central arm particle ID
  detectors}},\ }\href@noop {} {\bibfield  {journal} {\bibinfo  {journal}
  {Nucl. Instrum. Methods Phys. Res., Sec. A}\ }\textbf {\bibinfo {volume}
  {499}},\ \bibinfo {pages} {508} (\bibinfo {year} {2003})}\BibitemShut
  {NoStop}%
\bibitem [{\citenamefont {Adler}\ \emph
  {et~al.}(2004{\natexlab{b}})\citenamefont {Adler} \emph
  {et~al.}}]{CentralityIdentification}%
  \BibitemOpen
  \bibfield  {author} {\bibinfo {author} {\bibfnamefont {S.~S.}\ \bibnamefont
  {Adler}} \emph {et~al.} (\bibinfo {collaboration} {PHENIX Collaboration}),\
  }\bibfield  {title} {\bibinfo {title} {{High $p_{T}$ charged hadron
  suppression in Au + Au collisions at $\sqrt{s}_{NN}=200$ GeV}},\ }\href@noop
  {} {\bibfield  {journal} {\bibinfo  {journal} {Phys. Rev. C}\ }\textbf
  {\bibinfo {volume} {69}},\ \bibinfo {pages} {034910} (\bibinfo {year}
  {2004}{\natexlab{b}})}\BibitemShut {NoStop}%
\bibitem [{\citenamefont {Adler}\ \emph {et~al.}(2003)\citenamefont {Adler}
  \emph {et~al.}}]{PISA}%
  \BibitemOpen
  \bibfield  {author} {\bibinfo {author} {\bibfnamefont {S.}~\bibnamefont
  {Adler}} \emph {et~al.} (\bibinfo {collaboration} {PHENIX Collaboration}),\
  }\bibfield  {title} {\bibinfo {title} {{PHENIX on-line and off-line
  computing}},\ }\href@noop {} {\bibfield  {journal} {\bibinfo  {journal}
  {Nucl. Instrum. Methods Phys. Res., Sec. A}\ }\textbf {\bibinfo {volume}
  {499}},\ \bibinfo {pages} {593} (\bibinfo {year} {2003})}\BibitemShut
  {NoStop}%
\bibitem [{\citenamefont {Brun}\ \emph {et~al.}(1994)\citenamefont {Brun},
  \citenamefont {Bruyant}, \citenamefont {Carminati}, \citenamefont {Giani},
  \citenamefont {Maire}, \citenamefont {McPherson}, \citenamefont {Patrick},\
  and\ \citenamefont {Urban}}]{GEANT}%
  \BibitemOpen
  \bibfield  {author} {\bibinfo {author} {\bibfnamefont {R.}~\bibnamefont
  {Brun}}, \bibinfo {author} {\bibfnamefont {F.}~\bibnamefont {Bruyant}},
  \bibinfo {author} {\bibfnamefont {F.}~\bibnamefont {Carminati}}, \bibinfo
  {author} {\bibfnamefont {S.}~\bibnamefont {Giani}}, \bibinfo {author}
  {\bibfnamefont {M.}~\bibnamefont {Maire}}, \bibinfo {author} {\bibfnamefont
  {A.}~\bibnamefont {McPherson}}, \bibinfo {author} {\bibfnamefont
  {G.}~\bibnamefont {Patrick}},\ and\ \bibinfo {author} {\bibfnamefont
  {L.}~\bibnamefont {Urban}},\ }\href@noop {} {\bibinfo {title} {Geant detector
  description and simulation tool}} (\bibinfo {year} {1994}),\ \bibinfo {note}
  {{CERN-W5013, CERN-W-5013, W5013, W-5013}}\BibitemShut {NoStop}%
\bibitem [{\citenamefont {Adare}\ \emph {et~al.}(2014)\citenamefont {Adare}
  \emph {et~al.}}]{BiasFactor}%
  \BibitemOpen
  \bibfield  {author} {\bibinfo {author} {\bibfnamefont {A.}~\bibnamefont
  {Adare}} \emph {et~al.},\ }\bibfield  {title} {\bibinfo {title} {{Centrality
  categorization for {$R_{p(d)+A}$} in high-energy collisions}},\ }\href@noop
  {} {\bibfield  {journal} {\bibinfo  {journal} {Phys. Rev. C}\ }\textbf
  {\bibinfo {volume} {90}},\ \bibinfo {pages} {034902} (\bibinfo {year}
  {2014})}\BibitemShut {NoStop}%
\bibitem [{\citenamefont {Schnedermann}\ \emph {et~al.}(1993)\citenamefont
  {Schnedermann}, \citenamefont {Sollfrank},\ and\ \citenamefont
  {Heinz}}]{Thermal1}%
  \BibitemOpen
  \bibfield  {author} {\bibinfo {author} {\bibfnamefont {E.}~\bibnamefont
  {Schnedermann}}, \bibinfo {author} {\bibfnamefont {J.}~\bibnamefont
  {Sollfrank}},\ and\ \bibinfo {author} {\bibfnamefont {U.}~\bibnamefont
  {Heinz}},\ }\bibfield  {title} {\bibinfo {title} {{Thermal phenomenology of
  hadrons from 200A GeV S+S collisions}},\ }\href@noop {} {\bibfield  {journal}
  {\bibinfo  {journal} {Phys. Rev. C}\ }\textbf {\bibinfo {volume} {48}},\
  \bibinfo {pages} {2462} (\bibinfo {year} {1993})}\BibitemShut {NoStop}%
\bibitem [{\citenamefont {Greco}\ \emph
  {et~al.}(2003{\natexlab{b}})\citenamefont {Greco}, \citenamefont {Ko},\ and\
  \citenamefont {L\'evai}}]{Thermal2}%
  \BibitemOpen
  \bibfield  {author} {\bibinfo {author} {\bibfnamefont {V.}~\bibnamefont
  {Greco}}, \bibinfo {author} {\bibfnamefont {C.~M.}\ \bibnamefont {Ko}},\ and\
  \bibinfo {author} {\bibfnamefont {P.}~\bibnamefont {L\'evai}},\ }\bibfield
  {title} {\bibinfo {title} {{Partonic coalescence in relativistic heavy ion
  collisions}},\ }\href@noop {} {\bibfield  {journal} {\bibinfo  {journal}
  {Phys. Rev. C}\ }\textbf {\bibinfo {volume} {68}},\ \bibinfo {pages} {034904}
  (\bibinfo {year} {2003}{\natexlab{b}})}\BibitemShut {NoStop}%
\bibitem [{\citenamefont {Adare}\ \emph {et~al.}(2011)\citenamefont {Adare}
  \emph {et~al.}}]{CH_pp}%
  \BibitemOpen
  \bibfield  {author} {\bibinfo {author} {\bibfnamefont {A.}~\bibnamefont
  {Adare}} \emph {et~al.} (\bibinfo {collaboration} {PHENIX Collaboration}),\
  }\bibfield  {title} {\bibinfo {title} {{Identified charged hadron production
  in $p$$+$$p$ collisions at $\sqrt{s}=200$ and 62.4 GeV}},\ }\href@noop {}
  {\bibfield  {journal} {\bibinfo  {journal} {Phys. Rev. C}\ }\textbf {\bibinfo
  {volume} {83}},\ \bibinfo {pages} {064903} (\bibinfo {year}
  {2011})}\BibitemShut {NoStop}%
\bibitem [{\citenamefont {Aidala}\ \emph {et~al.}(2018)\citenamefont {Aidala}
  \emph {et~al.}}]{pi0Eta_CuAu}%
  \BibitemOpen
  \bibfield  {author} {\bibinfo {author} {\bibfnamefont {C.}~\bibnamefont
  {Aidala}} \emph {et~al.} (\bibinfo {collaboration} {PHENIX Collaboration}),\
  }\bibfield  {title} {\bibinfo {title} {{Production of $\pi^0$ and $\eta$
  mesons in Cu$+$Au collisions at $\sqrt{s_{_{NN}}}=200$ GeV}},\ }\href@noop {}
  {\bibfield  {journal} {\bibinfo  {journal} {Phys. Rev. C}\ }\textbf {\bibinfo
  {volume} {98}},\ \bibinfo {pages} {054903} (\bibinfo {year}
  {2018})}\BibitemShut {NoStop}%
\bibitem [{\citenamefont {Acharya}\ \emph {et~al.}(2020)\citenamefont {Acharya}
  \emph {et~al.}}]{pi0Eta_UU}%
  \BibitemOpen
  \bibfield  {author} {\bibinfo {author} {\bibfnamefont {U.}~\bibnamefont
  {Acharya}} \emph {et~al.} (\bibinfo {collaboration} {PHENIX Collaboration}),\
  }\bibfield  {title} {\bibinfo {title} {{Production of $\pi^0$ and $\eta$
  mesons in U+U collisions at $\sqrt{s_{_{NN}}}=$192 GeV}},\ }\href@noop {}
  {\bibfield  {journal} {\bibinfo  {journal} {Phys. Rev. C}\ }\textbf {\bibinfo
  {volume} {102}},\ \bibinfo {pages} {064905} (\bibinfo {year}
  {2020})}\BibitemShut {NoStop}%
\bibitem [{\citenamefont {Acharya}\ \emph
  {et~al.}(2022{\natexlab{a}})\citenamefont {Acharya} \emph
  {et~al.}}]{phi_SmallSysts}%
  \BibitemOpen
  \bibfield  {author} {\bibinfo {author} {\bibfnamefont {U.}~\bibnamefont
  {Acharya}} \emph {et~al.} (\bibinfo {collaboration} {PHENIX Collaboration}),\
  }\bibfield  {title} {\bibinfo {title} {{Study of $\phi$-meson production in
  $p$$+$Al, $p$$+$Au, $d$$+$Au, and $^3$He$+$Au collisions at
  $\sqrt{s_{_{NN}}}=200$ GeV}},\ }\href@noop {} {\bibfield  {journal} {\bibinfo
   {journal} {Phys. Rev. C}\ }\textbf {\bibinfo {volume} {106}},\ \bibinfo
  {pages} {014908} (\bibinfo {year} {2022}{\natexlab{a}})}\BibitemShut
  {NoStop}%
\bibitem [{\citenamefont {Acharya}\ \emph
  {et~al.}(2022{\natexlab{b}})\citenamefont {Acharya} \emph
  {et~al.}}]{pi0_smallSysts}%
  \BibitemOpen
  \bibfield  {author} {\bibinfo {author} {\bibfnamefont {U.}~\bibnamefont
  {Acharya}} \emph {et~al.} (\bibinfo {collaboration} {PHENIX Collaboration}),\
  }\bibfield  {title} {\bibinfo {title} {{Systematic study of nuclear effects
  in $p$ $+$Al, $p$ $+$Au, $d$ $+$Au, and $^{3}$He$+$Au collisions at
  $\sqrt{s_{_{NN}}}=200$ GeV using $\pi^0$ production}},\ }\href@noop {}
  {\bibfield  {journal} {\bibinfo  {journal} {Phys. Rev. C}\ }\textbf {\bibinfo
  {volume} {105}},\ \bibinfo {pages} {064902} (\bibinfo {year}
  {2022}{\natexlab{b}})}\BibitemShut {NoStop}%
\bibitem [{\citenamefont {Acharya}\ \emph
  {et~al.}(2022{\natexlab{c}})\citenamefont {Acharya} \emph
  {et~al.}}]{psi_SmallSyst}%
  \BibitemOpen
  \bibfield  {author} {\bibinfo {author} {\bibfnamefont {U.}~\bibnamefont
  {Acharya}} \emph {et~al.} (\bibinfo {collaboration} {PHENIX Collaboration}),\
  }\bibfield  {title} {\bibinfo {title} {{Measurement of $\psi(2S)$ nuclear
  modification at backward and forward rapidity in $p$$+$$p$, $p$$+$Al, and
  $p$$+$Au collisions at $\sqrt{s_{_{NN}}}=200$ GeV}},\ }\href@noop {}
  {\bibfield  {journal} {\bibinfo  {journal} {Phys. Rev. C}\ }\textbf {\bibinfo
  {volume} {105}},\ \bibinfo {pages} {064912} (\bibinfo {year}
  {2022}{\natexlab{c}})}\BibitemShut {NoStop}%
\end{thebibliography}

%
 
\end{document}